\documentclass[useAMS,usenatbib]{mn2e}
\usepackage{epsfig}
\usepackage{adjustbox}
\usepackage{amsmath}
\usepackage{graphicx}
\usepackage{subfig}

\def\simgt{\lower.5ex\hbox{$\; \buildrel > \over \sim \;$}}
\def\simlt{\lower.5ex\hbox{$\; \buildrel < \over \sim \;$}}

\title[AGN-stimulated Cooling of Hot Gas in Ellipticals]{AGN-stimulated Cooling of Hot Gas in Elliptical Galaxies}
 
\author[M. Valentini \& F. Brighenti]
{Milena Valentini$^{1,2}$\thanks{E-mail:
milena.valentini@sissa.it} and
Fabrizio Brighenti$^{2}$\thanks{E-mail:
fabrizio.brighenti@unibo.it}\\
$^{1}$SISSA, via Bonomea 265, I-34136 Trieste, Italy\\
$^{2}$Dipartimento di Fisica e Astronomia, Universit\`a di Bologna, via Ranzani 1, 
40126 Bologna, Italy}

\begin{document}

\date{Accepted 2015 January 13. Received 2015 January 13; in original form 2014 November 25}

\pagerange{\pageref{firstpage}--\pageref{lastpage}} \pubyear{2002}

\maketitle

\label{firstpage}

\begin{abstract}

We study the impact of relatively weak AGN feedback on the
interstellar medium of intermediate and massive elliptical
galaxies. We find that the AGN activity, while globally heating the ISM,
naturally stimulates some degree of hot gas cooling on scales of several kpc.
This process generates the persistent presence of a cold ISM phase, with
mass ranging between $10^4$ and $\ga 5 \times 10^7$ M$_\odot$, where
the latter value is appropriate for group centered, massive
galaxies. Widespread cooling occurs where the ratio of cooling to
free-fall time before the activation of the AGN feedback satisfies
$t_{\rm cool}/t_{\rm ff} \la 70$, that is we find a less restrictive threshold
than commonly quoted in the literature.
This process helps explaining the body of observations of cold gas
(both ionized and neutral/molecular) in Ellipticals
and, perhaps, the residual star formation detected in many early-type
galaxies. The amount and distribution of the off-center cold gas
vary irregularly with time.

The cold ISM velocity field is irregular, initially sharing the
(outflowing) turbulent hot gas motion.
Typical velocity dispersions of the cold gas lie in
the range $100-200$ km s$^{-1}$. Freshly generated cold gas often
forms a cold outflow and can appear kinematically misaligned with
respect to the stars.

We also follow the dust evolution in the hot and cold gas. We find that
the internally generated cold ISM has a very low dust content, with
representative values of the dust-to-gas ratio of $10^{-4} - 10^{-5}$.
Therefore, this cold gas can escape detection in the traditional
dust-absorption maps.

\end{abstract}


\section{Introduction} 
 \label{sec:introduction}

 Elliptical galaxies generally have a complex multiphase interstellar
 medium (ISM), which is a key component of the galactic ecosystem. The
 mixture of the various gas phases is usually different in early-type
 galaxies (ETGs) with respect to late-type systems and this, in turn,
 reflects on the different evolutionary pattern of morphological diverse
 galaxies. Reviews of hot, warm and cold gas in ETGs can
 be found in e.g. \citet[][]{caon2000, mb03araa, sarzi2006,sarzi2013,
 mulchaey2010, ellis2006, young2011, davis2011, serra2012}. The
cold ISM is often spatially extended, with irregular distribution and
kinematics \citep{caon2000}.
Dust is also present in all the phases of the ISM \citep[][]{temi2004,
temi2009, smith2012}, and the dust-to-gas ratio can be used to track
down the origin of the cold gas (see below).

A multiphase ISM is expected to keep the galaxy ``alive'', allowing
for (perhaps tiny) episodes of star formation (SF). Indeed, a sizeable
fraction of ETGs shows evidence of recent SF \citep{trager2000,kaviraj2007}.
Therefore, a grasp on the origin and evolution of the multiphase ISM would reveal vital
information on the evolution of ETGs, such as star formation history,
quenching mechanism(s), the
role of active galactic nuclei (AGN) feedback, black hole growth,
accretion of baryons from the environment and more.

In most ETGs a significant fraction of cold gas is thought to have an
external origin, as indicated by the misaligned kinematics of the cold
gas (both ionized and neutral) with respect to the stars \citep[e.g.][and references
  therein]{davis2011}. Cold gas is indeed commonly found around and
  bound to ETGs \citep{thom2012}. The fate of this cold gas reservoir
  is still unclear, but it is likely that some of this low entropy gas is falling onto the
  galaxy and reaches the core in $\approx 1$ Gyr. However,
  $\Lambda$CDM hydrodynamical simulations on the kinematics of accreted HI in ETGs indicate
  significant disagreement with respect to the observations, making
  the source of misaligned cold gas still an open question
  \citep{serra2014}. 

 Some of the cold ISM certainly comes from internal processes,
such as stellar mass loss or hot gas cooling
\citep{davis2011,werner2014,david2014}. 
Theoretical investigations about the formation of a
cold ISM phase in ETGs include
\citet{bm02,mb03,temi2007,parriott2008,McCourt12,gasp12,gaspari2012b,li2014};
Brighenti, Mathews \& Temi (2015). 
For massive galaxies, especially when located in high density environments, the
internally produced cold gas may well dominate over the accreted one.
\citet{cavagnolo2008} and \citet{voit2008} found that central galaxies in
clusters host a multiphase gas only when the central ICM (ISM) entropy
is below a threshold $kT/n^{2/3} \sim 30$ keV cm$^2$, providing a
clear link between the presence of cold gas and the properties of the
surrounding hot gas. In a recent theoretical investigation
\citet{lagos2014} argue that cooling from twisted hot gas halos results
in a large fraction ($\sim 46$ \%) of ETGs with internally generated misaligned cold gas,
explaining the ATLAS$^{\rm 3D}$ observations \citep{davis2011}.  

In this paper we study the generation of a spatially extended cold gas
phase in
ETGs by radiative cooling of the hot halo gas. The key question is
therefore if, and under which circumstances, the hot ISM can cool at
large distance from the galaxy center.
Multiwavelength observations often
show a strong spatial correlation between X-ray and H$\alpha$ images
\citep[e.g.][]{trinchieri1997,goudfrooij1998,trinchieri2002,temi2007,gastaldello2009,werner2014},
implying rapid local cooling. Furthermore, the connection between
cold gas, radio emission and AGN-driven X-ray disturbances \citep[see
references above, also][]{blanton2011} suggests the key point
that spatially extended cooling has been triggered by the AGN activity. Evidently, while
AGN feedback suppresses the total cooling rate with respect to
standard cooling flow models \citep[e.g.][]{mcnamara2007}, it also stimulates some gas condensation on
scales of several kpc.

Guided by these
observational results, we present here a detailed investigation of
spatially extended gas cooling resulting from AGN feedback. We do not
address the long term fate of the cooled gas --- it can be in part accreted onto
the central black hole \citep{gaspari2012b}, it can settle in a rotating disk
\citep{brighenti1996,brighenti1997} or it may be re-heated to the hot
gas temperature by thermal conduction or other processes.

For the reasons outlined in \citet{gasp12} we consider collimated,
non-relativistic outflows as the dominant feedback mechanism in local
ETGs. The AGN feedback process is far more complicated than we
consider here and is not appropriately understood 
\citep[see, e.g.][for a discussion on the numerical limitations in simulating
  AGN heating in galaxies and clusters]{gasp12}.
However, for the objectives of this paper it is not
necessary to accurately model the whole feedback cycle. An idealized
feedback scheme as the one described in the next section is adequate
enough to demonstrate the physical existence and the basic properties of widespread gas
cooling in ETGs.

The thermal stability of gas in galactic or cluster cooling flows has been investigated
since the discovery of the hot ISM and ICM, through both linear
stability analysis and numerical simulations \citep[see, for
  instance][]{mathews1978,cowie1980,bodo1987,malagoli1987,balbus1988,
loewenstein1989,balbus1989,malagoli1990,hattori1990,reale1991,binney09,
McCourt12,Joung2012,gaspari2012b,li2014}.
However, linear stability analysis provides an incomplete understanding when
strong density perturbations are present. It is quite obvious that an
overdense clump of gas can cool fast and separate from the hot
phase. The density contrast threshold for this localized cooling
depends on the properties of the background medium and the
perturbation \citep[see][for a recent numerical
investigation]{Joung2012}. Therefore, the key astrophysical problem is
to understand if such strong perturbations are naturally generated
inside elliptical galaxies. As we discuss below, the answer delivered
by our simulations is yes.
We prefer to keep the
assumptions of our models to a minimum. We do not use an {\it ad hoc}
distributed heating to ensure global thermal equilibrium, nor we
superimpose an artificial turbulent velocity field to generate density
perturbations. Adopting low-power, collimated outflows as the only 
perturbing means of the ISM, we aim to probe the importance of
AGN-stimulated cooling in ETGs in a simple yet solid way.

\section{Numerical Procedure}
\label{sec:models}

The simulations have been carried out with a modified version of the
eulerian code ZEUS-2D \citep{zeus92}, developed to include in the
set of hydrodynamical equations source and sink terms describing
galactic scale cooling flows.
These terms include mass and energy injection by the stellar
population, as described in \citet{mb03araa}. The present time type Ia
supernova rate adopted here is 0.02 SNu\footnote{These terms have no
  significant impact on our results and have been included just for
  consistency with previous cooling flow calculations.}.
The code has been further enhanced to deal with a second fluid
representing the dust. Dust density $\rho_{d}$
obeys to the equation:
\begin{equation}
\frac{\partial\:\rho_{d}}{\partial t}\:+\: \nabla \cdot (\rho_{\:d} \: 
\mathbf{u}) \:=\: \alpha_{\ast}\rho_{\ast}\:\delta_{\ast}\: - 
\:\dot{\rho}_{sputt} + \dot{\rho}_{growth}\:,
\label{eq:1}
\end{equation}

\noindent
where $\mathbf{u}$ is the velocity of the gas, $\alpha_{\ast}$ the
specific rate of stellar mass loss, $\rho_{\ast}$ the stellar density and a uniform stellar dust to gas
mass ratio $\delta_{\ast}=0.01$ has been assumed.
On the right hand side of equation $\ref{eq:1}$, besides the stellar source term for dust, there are the sink term due
to grain sputtering and the
source term due to grain growth, $\dot{\rho}_{sputt}$ and $\dot{\rho}_{growth}$, respectively. 

Thermal sputtering \citep{DS1979, tsai1995} rate is estimated as: 
\begin{equation}
\dot{\rho}_{\rm sputt} \,=\, \frac{\rho_{d}}{\tau_{\rm sputt}} \,=\,
n_d \: \langle\dot{m}_{\rm gr}\rangle \:,
\label{eq:2}
\end{equation}

\noindent
where $n_d \,=\, \rho_d/\langle m_{\rm gr} \rangle$ 
is the number density of dust grains,
$\langle\dot{m}_{\rm gr}\rangle$ is the average mass sputtering rate per grain,
$\dot{m}_{\rm gr}=4 \pi a^2 \rho_{\rm gr} \dot{a}\:$ being the mass sputtering
rate for a grain of given radius $a$. Here, $\rho_{\rm gr}=3.3$ g cm$^{-3}$
is the density of silicate grains \citep{temi2003}, $\dot{a}$ is the rate
at which the grain radius decreases \citep{tsai1995, mb03} and the
resulting average sputtering time is:
\begin{equation}
\begin{split}
\tau_{\rm sputt} &=\frac{\langle m_{\rm gr} \rangle}{\langle\dot{m}_{\rm gr}\rangle}=\frac{\int_{a_{min}}^{a_{max}} \, a^{3-s} \:
\mathrm{d}a}{3|\dot{a}| \int_{a_{min}}^{a_{max}} \, a^{2-s} \:
\mathrm{d}a} = \frac{\langle a^3 \rangle}{3|\dot{a}| \langle a^2
  \rangle}\\
&= \frac{0.03}{n_p} \left [ 1+\left ( \frac{2\times
    10^6}{T} \right )^{2.5} \right ] \;\; {\rm Myr}\:.
\end{split}
\label{eq:3}
\end{equation}

\noindent
In equation \ref{eq:3}, $s=3.5$ \citep{MNR1977}, $a_{min}=0.01$
$\mu$m and $a_{max}=1$ $\mu$m \footnote{We checked that these values of
minimum and maximum grain size do not affect our results.}. Finally, $\langle a^3
\rangle = \int{a^{3-s} da}/\int{a^{-s} da}$ and similarly $\langle
a^2 \rangle$.

The grain growth rate in cold and dense gas 
\citep{dwek1988, Hira2011, Hira2012, Zhu2008}
is estimated as:
\begin{equation}
\dot{\rho}_{\rm growth} \:=\: n_d \: \langle\dot{m}_{\rm gr,growth}\rangle \:,
\label{eq:4}
\end{equation}

\noindent
where the growth rate of each grain of mass $m$ and radius $a$
is $\dot{m}_{\:gr,growth}=\pi a^2 f n_m m_m v_m$, $f=0.3$ being the
probability that a colliding atom/metal sticks on the grain surface,
$n_m$ and $v_m\propto \sqrt{T}$ \citep{dwek1998} the metal number
density and thermal velocity (assuming an atomic weight $m_m/m_p = 20$). 
The averaged dust density growth rate is therefore: 
\begin{equation}
\dot{\rho}_{\rm growth} = \rho_d \frac{\langle \dot{m}_{\rm gr,growth}\rangle}{\langle m_{\rm gr}\rangle}
 = \frac{3}{4}\frac{\rho \rho_{\rm
    d}}{\rho_{\rm gr}} f v_m (0.01 - \delta) \frac{\langle a^2
  \rangle}{\langle a^3 \rangle}  \:,
\end{equation}

\noindent
where $\rho$ is the gas density and it has been taken into account that
when the dust to gas mass ratio $\delta$ approaches the value $0.01$
all the relevant metals are locked in dust grains. 

Radiative cooling describing the loss of energy for X-ray emission
has been included with the term $-n_i n_e \Lambda(T)$
in the thermal energy equation, where the cooling function $\Lambda$
is calculated according to \cite{sutherland93}, assuming solar metallicity.
The cooling has been extended to temperatures lower than $T=10^4$ K
following \cite{dalgarno72} and assuming a value $x=10^{-2}$ of the
fractional ionization. The cooling is truncated at $T=50$ K. Dust-induced
cooling, important for $T\ga 10^6$ K for a dust-to-gas ratio $\sim
10^{-2}$, has been taken into account according to \citet{mb03}.
We are well aware that the cooling and chemistry of cold gas is
extremely complicated and our treatment for the $T<10^4$ K gas is
approximate.

The two-dimensional $z\times R$ computational domain is made up
of $1100\times700$ zones. The resolution is $\Delta z = \Delta R =5$ pc
up to $z \times R = 5\times3$ kpc. The outer
$100$ points in both dimensions are separated by increasing distances which are the
terms of a geometric progression whose common ratio is $1.095$.
This peculiarity of the grid allows to study the central region of the
computational domain, where most of the cooling process is expected
to occur, in an accurate way.

\subsection{Models}
\label{subsec:models}

We consider three models of elliptical galaxies: the {\it central
  galaxy} (thereafter {\it CG}), located at the center of an X-ray bright
group, the {\it isolated galaxy} ({\it IG}), an isolated, fairly massive
elliptical, and the {\it low mass galaxy} ({\it LM}), an intermediate
elliptical, which represents a more common member of the early-type galaxy
family. The key difference between the first and the two latter
models is the amount of hot gas within the galaxy, which in turn
reflects on a significant difference in the cooling time profile (see
Figure \ref{fig:ratios}). In the {\it CG} most of the hot gas within 10
kpc is expected to be intragroup medium (IGM) \citep[e.g.][]{bm99},
while for the {\it IG} and {\it LM} all the hot gas originates from stellar mass
loss \citep[]{math86, loew87, ciot91, mb03araa}.

\subsubsection{Central Galaxy}

\begin{figure}
\begin{center}
\adjincludegraphics[trim=1.2cm 4.5cm 0cm 5.4cm, clip, max width=0.71\textwidth]{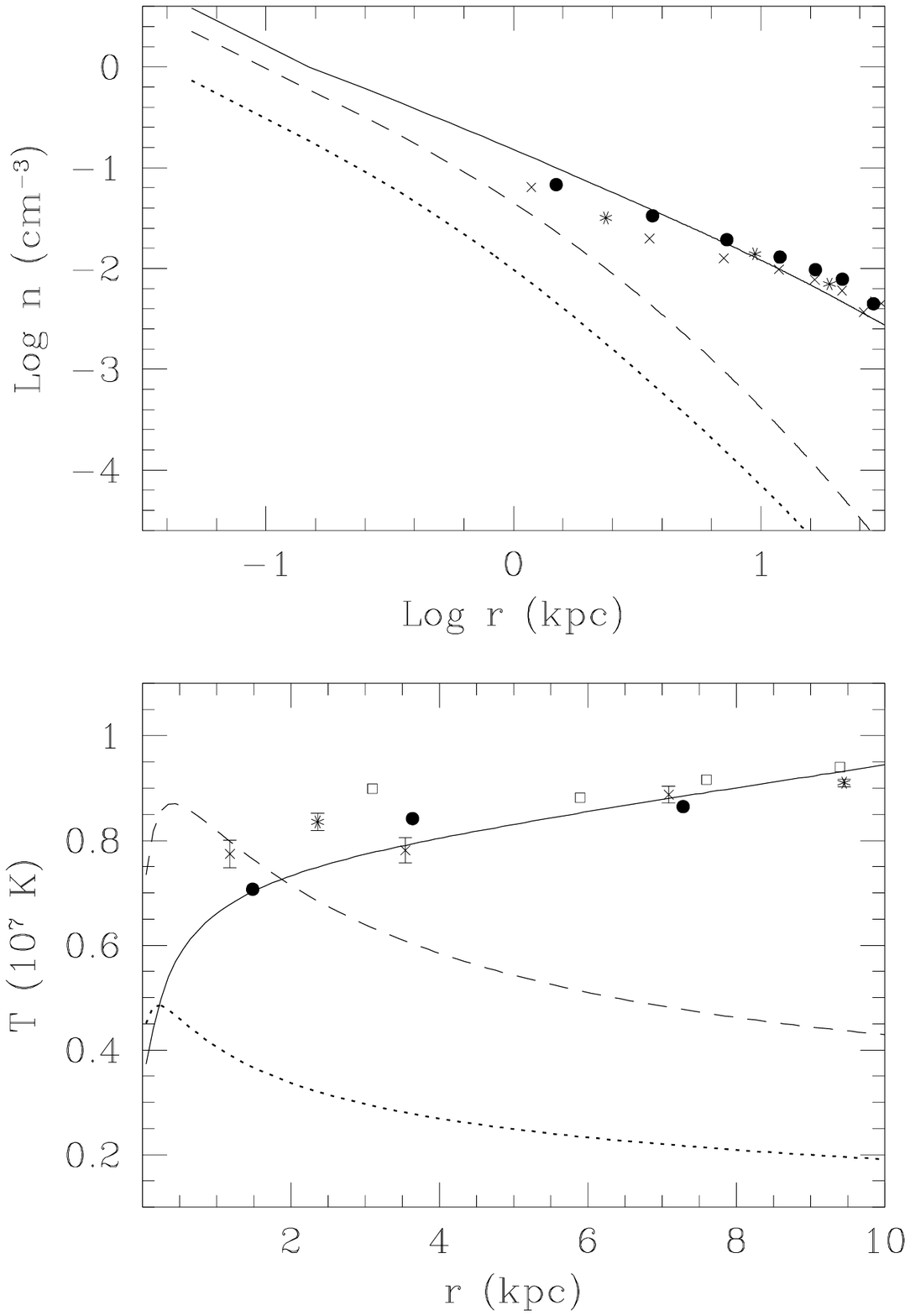} 
\end{center}
\caption{{\it Upper panel:} Density profile just before the outflow
  activation for the central
  galaxy (solid line), the isolated galaxy (dashed line) and the low mass galaxy (dotted line). The
  points represent data for NGC 5044 taken from \citet[]{buote03, buote04, david09}.
  {\it Bottom panel:} initial (3D) temperature profiles for the central
  galaxy (solid line), the isolated galaxy (dashed line) and the low mass galaxy (dotted line). Data
  points for NGC 5044 are taken from \citet[]{buote03, buote04, david09}.
}
\label{fig:initial}
\end{figure}

We model the {\it CG} to agree with NGC 5044, the brightest galaxy in the
homonymous group \citep[see][for further details]{gasp11b,gasp12}.
The gravitating mass consists of a NFW \citep[]{navarro96} halo with
virial mass $M_{\rm vir}=4\times 10^{13}$ M$_\odot$ and  concentration
$c=8.5$, plus a de Vaucouleurs profile \citep[]{mellier87}
with effective radius and stellar mass $r_{\rm e} = 10$ kpc and
$M_* = 3.5 \times 10^{11}$ M$_\odot$. 
Although inconsequential here, we also include a central
black hole with mass $M_{\rm BH}=4\times 10^8$ M$_\odot$. Regarding the
hot gas, we adopt the observed temperature profile for NGC 5044 
\citep[]{buote03, buote04, david09}
as initial temperature distribution, then we calculate the
gas density profile assuming hydrostatic equilibrium, choosing the
central density to approximately agree with observations. Before to
activate the AGN outflow, we let the system develop a classical
cooling flow, by evolving the flow for 100 Myr\footnote{This has a
  negligible  effect on the results presented in the following. We decided
  to kick off the AGN feedback from a pure cooling flow just to be
  consistent with the {\it IG} and {\it LM} models.}. During this time, the
initially present gas mixes with gas shed by the (old) stellar population
of the galaxy\footnote{We refer the reader to \citet[]{mb03araa} for the
hydrodynamical equations solved here and the relevant source terms for
mass and energy.}. 
The gas density and temperature profiles just before the activation 
of the AGN outflow are shown in Figure \ref{fig:initial}.
The hot gas mass within $r=10$ kpc is $\sim 2.6 \times 10^9$
M$_\odot$, and the bolometric X-ray luminosity within the virial
radius ($r_{\rm vir} \sim 900$ kpc) is $L_{\rm X} \sim 1.5\times
10^{43}$ erg s$^{-1}$. In Figure \ref{fig:lblx} it can be seen where
this galaxy model is located in the B-band luminosity versus
the bolometric X-ray luminosity plane for early type galaxies. As
expected, it
resides in the high luminosity tip of the observed $L_B - L_X$ relation
\citep{ellis2006}, populated by massive galaxies at the center of
groups or sub-clusters.

\subsubsection{Isolated Galaxy}

For this model, representative of massive but non central ellipticals,
the stellar profile follows a de Vaucouleurs distribution, with total
mass $M_* = 5 \times 10^{11}$ M$_\odot$ and effective radius $r_{\rm
  e} = 6$ kpc. The dark halo has a NFW profile with virial mass
$M_{\rm vir } = 10^{13}$ M$_\odot$ and concentration 10.0.

\begin{figure}
\begin{center}
\adjincludegraphics[trim=3.25cm 7.9cm 5.2cm 10.2cm, clip, max width=0.5\textwidth]{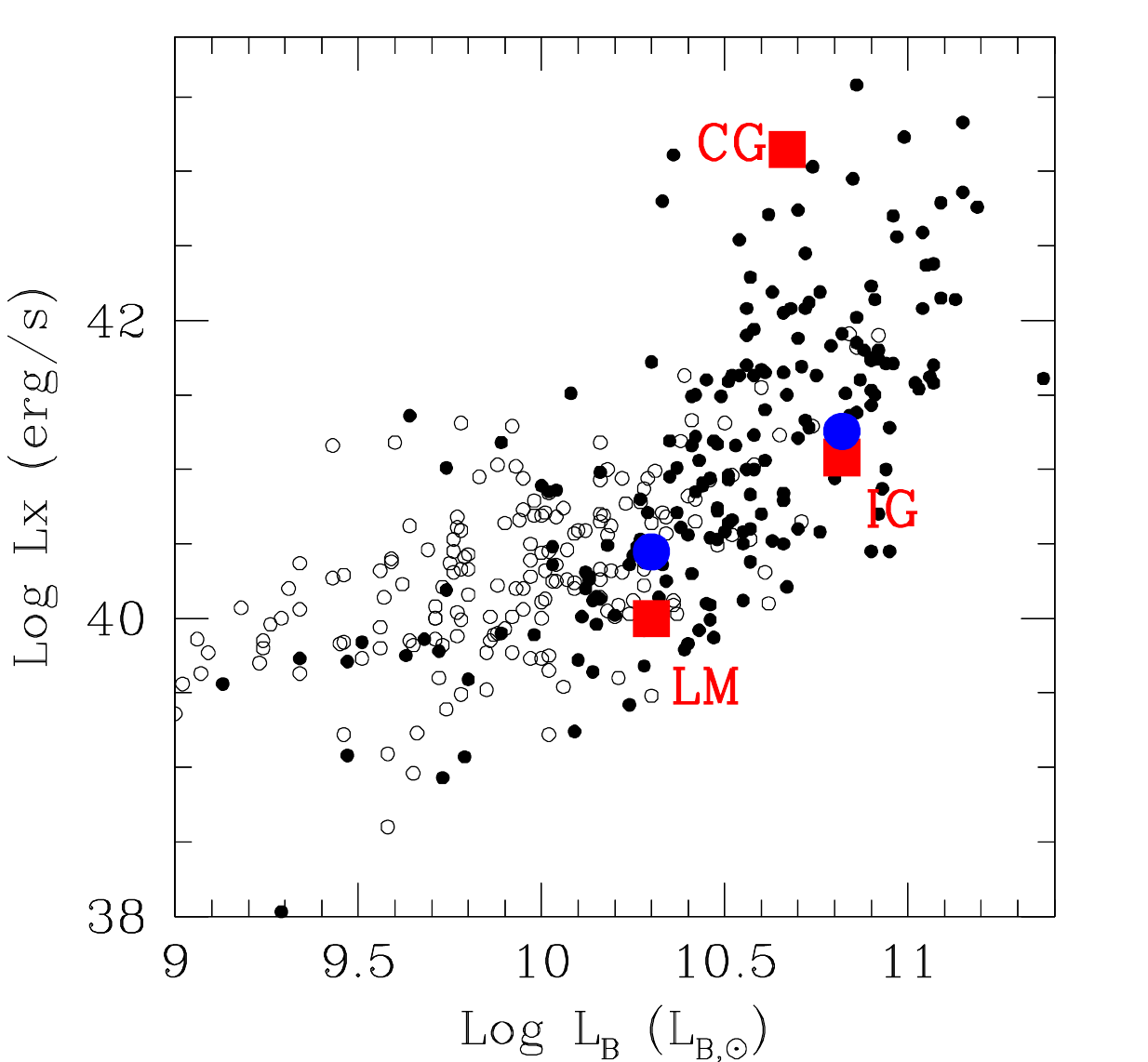} 
\end{center}
\caption{$L_B-L_X$ diagram for the elliptical galaxies sample of
  \citet[]{ellis2006}. Filled circle represent galaxies with detected
  X-ray emission; open circles are used for upper limits.
The location of our models, just before the AGN activation, is marked
with the red squares when only the X-ray luminosity of the hot gas is
considered. Big blue dots individuate the models when the contribution
of the low mass X-ray binaries is added \citep{kim2004}.
}
\label{fig:lblx}
\end{figure}

The initial ISM is generated by evolving a classic cooling flow for 1 Gyr
(the choice of this time is arbitrary and has no consequence on the
results, since the flow reaches a quasi-steady state after $\sim 100$ Myr).
The ISM entirely builds from the mass loss from the stellar population
\citep[e.g.][]{loew87, mb03araa}.
Just before the AGN outburst, 
the hot ISM mass within 10 kpc is $\sim 1.9 \times 10^8$ M$_\odot$,
the X-ray bolometric luminosity being $L_x\sim 1.2\times 10^{41}$ erg
s$^{-1}$. Figure \ref{fig:lblx} shows that this model occupies an
intermediate and uninteresting place in the $L_B - L_X$ diagram.
The initial gas density and temperature profiles are shown in Figure
\ref{fig:initial} as dashed lines. Notice as pure cooling flows in
elliptical galaxies show an overall negative temperature gradient, because of
the central steep potential well and the resulting compressional
heating.

\subsubsection{Low Mass Galaxy}
This model is a scaled down version of the {\it IG} described above. The stellar
mass is now $M_* = 1.5 \times 10^{11}$ M$_\odot$ and the effective radius
$r_{\rm e} = 3.16$ kpc, a value which agrees with the magnitude-$r_{\rm e}$ 
relation \citep[]{faber97}. The NFW dark halo has a virial mass
$M_{\rm vir } = 1.5 \times 10^{12}$ M$_\odot$ and concentration 11.9.
The mass of hot gas within $r<10$ kpc is only
$\sim 3.4 \times 10^7$ M$_\odot$ ($\sim 1.15\times 10^7$ M$_\odot$
within $r_{\rm e}$), with a X-ray (bolometric) luminosity
$L_x\sim 10^{40}$ erg s$^{-1}$, which places this model within the
lower envelope of the $L_B - L_X$ relation.

As common in intermediate and low mass ellipticals, only the gas in
the inner region is inflowing, while the ISM at larger radii forms a
subsonic outflow, powered by the SNIa \citep[e.g.][]{mathews71, mb03araa}.
For the assumed SNIa rate (0.02 SNu), the transition between inflow and outflow
is located at $r\sim 2$ kpc. We note that for this relatively small
galaxy the properties of the ISM are quite sensitive to the SN heating.
Doubling the SNIa rate results in a global outflow (but the inner
$\sim 200$ pc), with velocity increasing almost linearly from $\sim 0$
at 200 pc to $\sim 60$ km s$^{-1}$ at 3 kpc. The hot gas mass (within
10 kpc) is
reduced to $1.9 \times 10^7$ M$_\odot$, while the X-ray luminosity drops to
a very small value of $1.8 \times  10^{39}$ erg s$^{-1}$ (see
\citet{mb03araa} for more details on the effect of the SNIa rate on
the gas flow in elliptical galaxies). 

\subsection{Modeling the outflow}
\label{subsec:outflow}

\subsubsection{Single event}
\label{subsec:single event}

A subrelativistic and massive outflow \citep{brighenti2006,gasp12} which lasts $2$ Myr
\citep[e.g.][]{fabian2012} is our single event AGN mechanical outburst.
The outflow energy injection is simulated dispensing 
velocity to the gas located in a box of
$(R \times z) = (100 \times 100)$ pc. This {\it active region} is the inner
corner of the computational domain, near the origin of the R and
z-axis\footnote{Numerical experiments show that injecting the kinetic energy from a
nozzle along the inner boundary conditions at $z=0$ does not change the results.}.

The number of cells by which resolve the interior of the jet has been chosen
arbitrarily, the reasonable assumption is that the physics of the outflow does not influence
the AGN feedback effects on kpc scale-length and the IGM global properties.

\begin{figure}
\begin{center}
\adjincludegraphics[trim=2.7cm 4.7cm 6.7cm 4.8cm, clip, angle=-90,max width=0.45\textwidth]{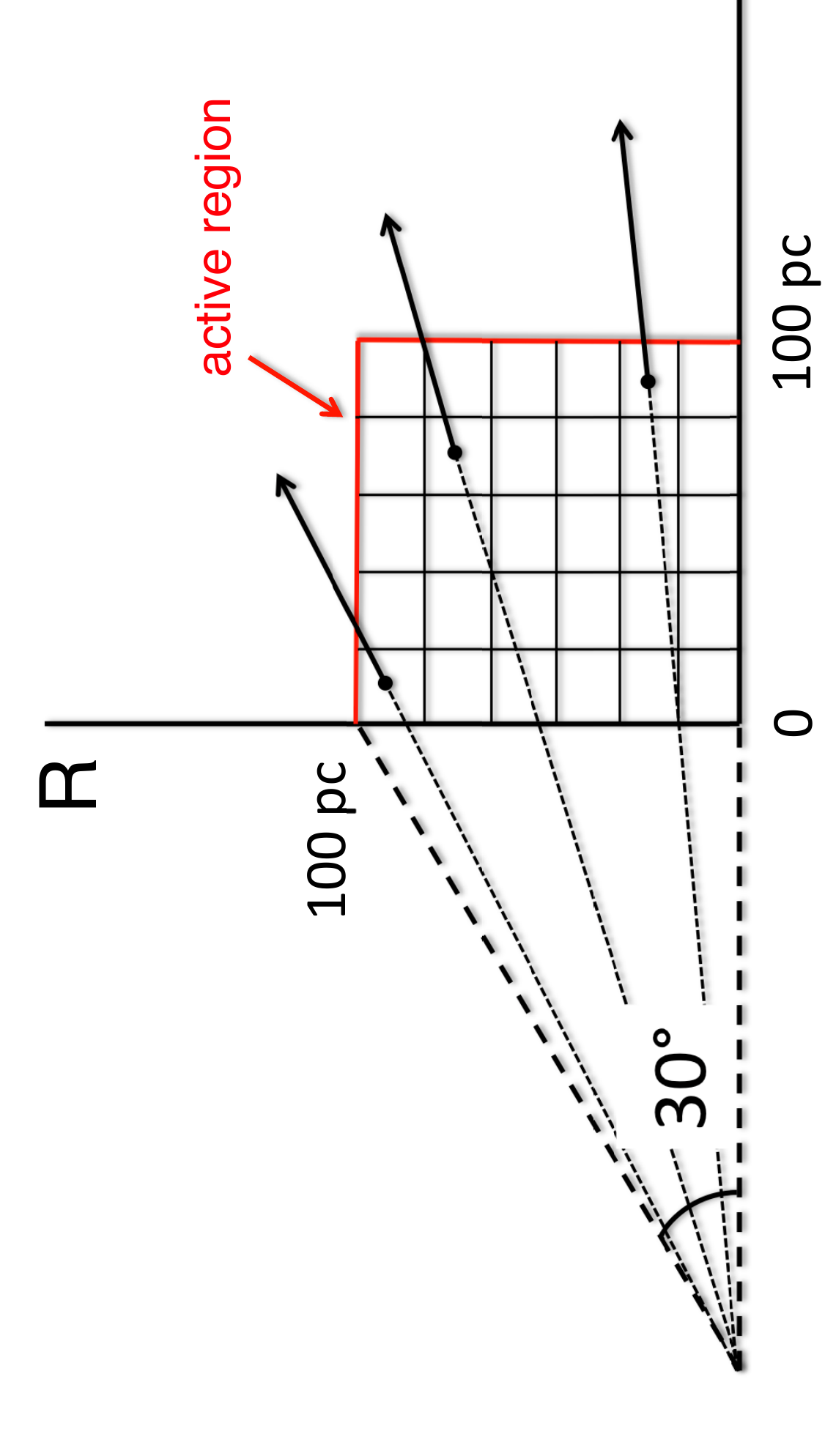} 
\end{center}
\caption{Scheme showing the method we chose to implement a conical
  outflow (see text). The inner corner of the 2D cylindrical grid, with the $100
  \times 100$ pc$^2$ active region highlighted, is represented. We
  note that in the code the $z$ and $R$ components of the velocity are
  located on different positions \citep[see][]{zeus92}. Here we show
  the velocity vectors originating from the grid center just for illustrative purposes.
}
\label{fig:conico}
\end{figure}

The outflow velocity is given to the cells of the active region while the AGN
is switched on, this energy input representing a spatially localized
source term in the hydrodynamical equation of momentum.
The velocity is given setting its module and defining a way for describing
the components along R and z axes. We simulate both cylindrical and
conical outflows, although we will focus mainly on the latter case.
Cylindrical outflows have been
simulated by equating the z-component of the velocity to the velocity
module and setting the R-component equal to zero in all the aforementioned
zones. On the other hand, for conical outflows we set the velocity
vectors as illustrated in Figure \ref{fig:conico}. At any given grid
point the velocity is radially directed, with the center located at
$(z,R)=(0,-100$ pc/tg(30$^{\circ}))$.  
The results are quite insensitive to the exact method of
implementing a conical outflow. 

All the models have been evolved for $50$ Myr, in order to investigate the short
term evolution of a single AGN outburst. Each performed simulation was given
a name: the identifying convention encodes the model of galaxy and both the
shape and the velocity in km/s of the outflow. For instance, model {\it IG4000} refers
to the conical outflow with $v_{jet}=4000$ km s$^{-1}$ activated in the isolated
galaxy.

\subsubsection{AGN feedback: recurrent outbursts}
\label{subsec:feedback}

In these models we consider the whole AGN feedback cycle, where the
gas cools and accretes on the central black hole, triggering an AGN
outburst which heats the ISM. When the gas cools in the central
region, the cycle is assumed to start again. Needless to say, the
physics of AGN feedback is very complex and poorly known. We have no
pretension to provide a self-consistent model here, we just need a
simple scheme to set off outflows intermittently, in order to mimic the
dynamical --- sometime violent --- evolution of the ISM in
elliptical galaxies. We thus employ a simplified version of the feedback
scheme used elsewhere
\citep[e.g.][]{brighenti2006,gasp12}.  Although this feedback is not
fully self-regulated, the modeled outflows reproduce reasonable
energies and powers (see Section \ref{sec:energetics}).

The adopted recipe has two main ingredients:
the way of numerically describing the outflow and its activation timing. 
We adopted a simple method in which the mechanical feedback activation
is triggered by the presence of cooling gas in
the innermost region ($r \le 100$ pc) of the simulated galaxy. This assumption
mimics the so-called cold accretion idea \citep[e.g.][and references
therein]{gaspari2012b}: AGN outflows are fed by accretion of gas
cooled out of the hot ISM, which accretes onto the central black hole.

After several numerical experiments, we decided to adopt a 
{\it dropout} term $- q(T) \rho /t_{cool}$, with
$q(T)=q_0\,exp[-(T/T_c)^2]$, \citep{bm02} in
the continuity equation to individuate and remove from the grid gas
currently cooling to low
temperatures. When gas dropout within $r\le 100$ pc occurs
(at least $1$ M$_\odot$ in one timestep), the outflow is set off.
We choose $q_0=2$ and
$T_c=5\times 10^5$ K to describe $q(T)$, although the results are
insensitive to the exact values of these parameters.
This dropout method might result unpalatable and contrived, but has
been shown to provide realistic estimates for cooled gas masses,
limiting the effect of numerical overcooling over the traditional
approach (see \citet{bm02}, Brighenti, Mathews \& Temi (2015) and the
Appendix). Beside rectifying the numerical overcooling problem, there
is an even more crucial reason to prefer the dropout technique. In
order to calculate the genuine cooling rate at large distances from
the center, it is preferable to avoid the presence of cold gas in the
inner region, which would be ejected through an AGN outflow possibly
reaching the off-center region and mixing with the freshly cooled hot gas.

We note that we use the dropout term only near the center of the
system (for $r<500$ pc), so that the off-center gas cooling (see
below) is calculated in the standard way, that is letting the gas
cooling to low temperatures and keeping it on the numerical grid,
although this is know to cause spurious overcooling (see Appendix).

Once the outburst has been activated, the gas which resides in the inner
region defined by $R<100$, $z<100$ pc is given a constant
velocity as long as the condition on the presence of dropping out
gas in $r\le 100$ pc is met.

All the models consider a sequence of conical outflows, their opening angle
being $\theta=30^{\circ}$, and evolve up to $50$ Myr.
Models of recurrent feedback ({\it FB}) were given a name which shows the
model of galaxy and the velocity in km/s of the outflows, for instance {\it
 CG-FB2000}.

\section{Cooling and Timescales}
\label{subsec:timescales}

Here we briefly discuss the relevant timescales for thermal
instability for our three models. It is well known that linear
(infinitesimal) perturbations in the central
regions of ETGs and galaxy clusters are generally stable in the pure
hydrodynamical case
\citep[e.g.][]{malagoli87, balbus88, loew89, balbus89, malagoli90, binney09}. 
In fact, slightly overdense fluid elements oscillates at the
Brunt-V\"as\"ail\"a frequency $\omega_{\rm BV}=2\pi / \tau_{\rm BV}$,
where usually the oscillation period $\tau_{\rm BV} \ll t_{\rm
  cool}$. The oscillation center of the perturbation slowly sinks
toward the center of the system, in a time only slightly shorter than the flow time
of unperturbed gas.

\begin{figure*}
\begin{center}
\adjincludegraphics[trim=0.15cm 7.8cm 1.7cm 8.3cm, clip, max width=0.9\textwidth]{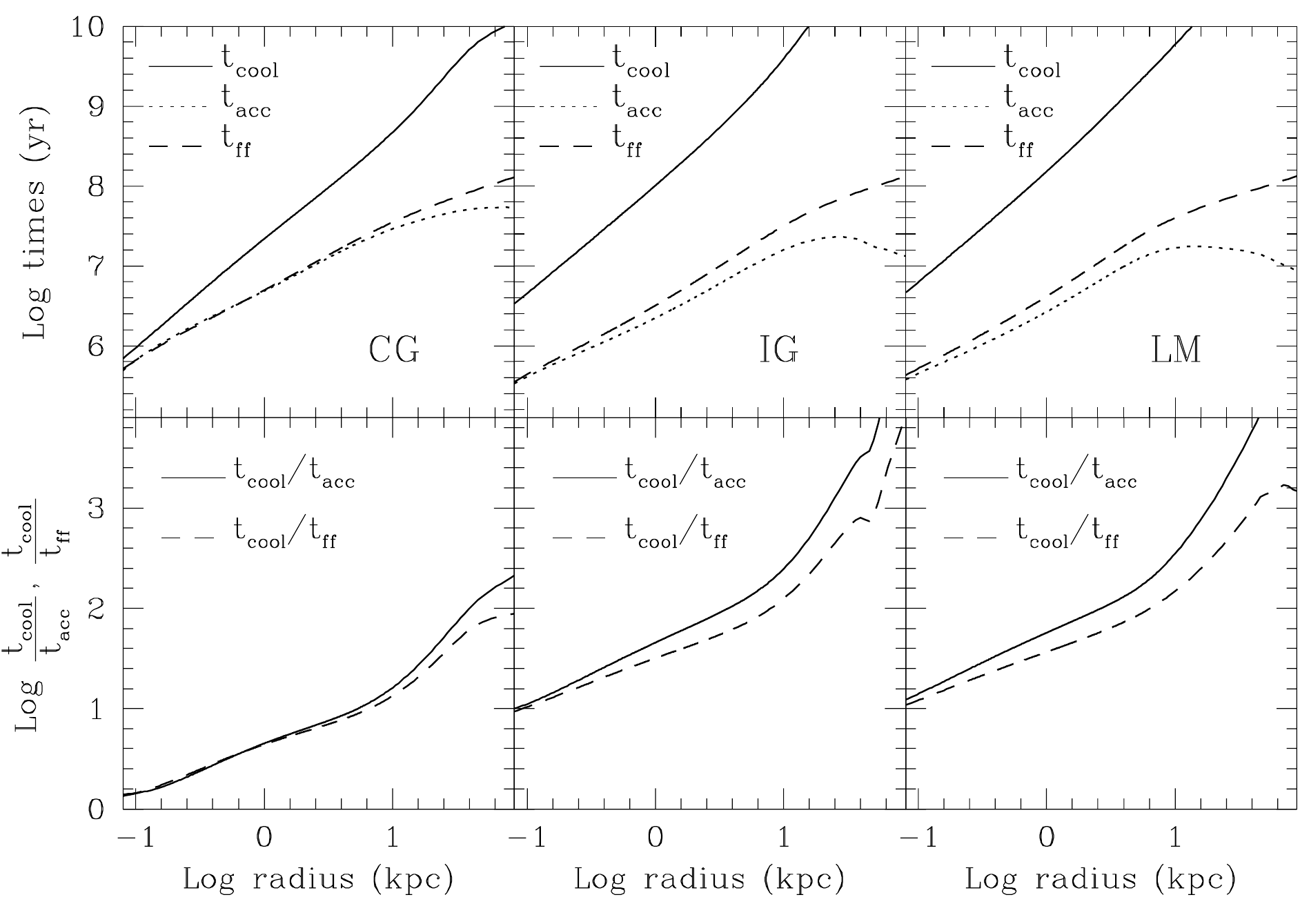} 
\end{center}
\caption{{\it Upper panels:} Cooling time, acceleration time and free fall time
  (see definitions in the text) just before the AGN outburst, for the {\it CG} (left), the {\it IG}
  (middle) and {\it LM} (right).
  {\it Bottom panels:} ratios of relevant timescales.
}
\label{fig:ratios}
\end{figure*}

However, the linear stability analysis is of limited use, as in
realistic galactic environments violent dynamical processes are
expected to generate perturbations of finite amplitude. Indeed, X-ray
images of cool cores clearly show large variations of gas density on
all scales probed by current telescopes.

\citet[]{bm02} have shown that spatially extended gas cooling in a
cool core can happen when the hot gas is perturbed by the action of
AGN feedback heating. \citet[]{bm02} found that localized
non-linear perturbations form in regions of converging flow and are
able to cool without undergoing oscillations.
Recently,
\citet[]{McCourt12, Sharma12, gaspari2012b, singh2015} have studied in detail the
cooling process in galaxy cluster cores perturbed by AGN feedback
\citep[see also][]{gasp11a}. They provide a heuristic criterion for the
onset of off-center thermal instability: the ICM undergoes widespread
cooling when stirred by AGN feedback if $t_{\rm cool}/t_{\rm ff} \la
10$, with the free fall time $t_{\rm ff}\sim t_{\rm dyn}=(2r/g)^{1/2}$, $g$ being the local
gravitational acceleration. \citet[]{Joung2012} investigated the
evolution of nonlinear perturbations in the Galactic halo and found
that for $t_{\rm cool}/t_{\rm acc} \la 1$ the ``cloud'' cools before
being disrupted by hydrodynamical instabilities. Here $t_{\rm
  acc}=c_s/g$, which is similar to $t_{\rm ff}$.

In Figure \ref{fig:ratios} we show the ratios $t_{\rm cool}/t_{\rm
  acc}$ and $t_{\rm cool}/t_{\rm ff}$ for our three systems, at the
time just before the AGN activation. For the {\it CG} model the instability
criterion $t_{\rm cool}/t_{\rm ff} \la 10$ is met for $r\la 6$ kpc,
while for {\it IG} and {\it LM} no unstable region exists outside the galactic
core $r\la 100$ pc, according to \citet[]{Sharma12}.

However, in the following we show that the commonly adopted instability
criterion is too restrictive. We find that extended cooling within elliptical
galaxies occurs for $t_{\rm
  cool}/t_{\rm ff} \la 70$ for realistic, recurrent, gentle AGN
outbursts. 
We note that the empirical entropy criterion for star formation and multiphase
gas in galaxy clusters cool cores, $K=kT/n^ {2/3} \la 30$ keV cm$^2$
\citep{voit2008,cavagnolo2008} describes more accurately the spatial
extent of the region where cooling happens. For the profiles
shown in Figure \ref{fig:initial} we find that the radius where $K=30$
keV cm$^2$ is 17, 5.4, 3.6 kpc for model {\it CG, IG} and {\it LM},
respectively. These numbers compare well with the maximum size of the
region where cooling occurs.

\section{Spatially extended cooling: single AGN outflow}
\label{sec:single}

\begin{figure*}
\begin{center}
\adjincludegraphics[trim=1.3cm 7.cm 2.8cm 5.9cm, clip, max width=0.95\textwidth]{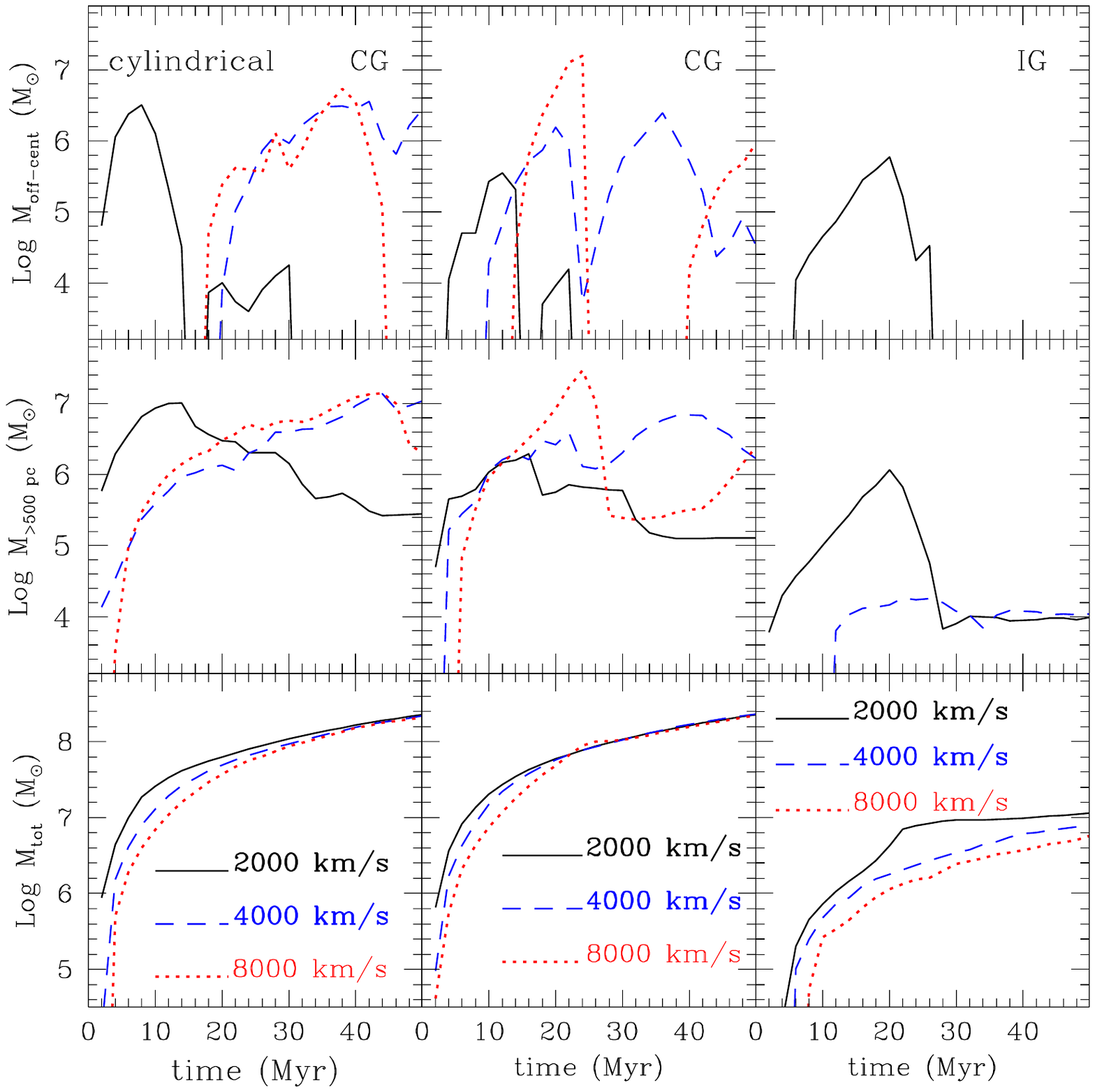} 
\end{center}
\caption{{\it Upper left panel:} Time evolution of the cold gas mass
  in the off-center region ($R>150$ pc, $z>500$ pc), for {\it CG} models
  with cylindrical jets {\it CGcyl2000}, {\it CGcyl4000} and {\it
    CGcyl8000}. Here and in the following, times are measured since the start of the AGN aoutflow. {\it
    Middle left panel:} as above for the total cold gas mass in the region
  $r>500$ pc, which includes the filament along the $z-$axis. {\it Lower left panel:} same for the total cold gas
  mass, most of which is located at the very center of the galaxy ($r
  < 300$ pc). {\it Central panels:} same as left panels but for the
  {\it CG} galaxy with conical jets (30 degrees) models {\it CG2000}, {\it CG4000} and {\it CG8000}. 
{\it Right panels:} same as central panels but for the
  isolated galaxy.  }
\label{fig:mcold_con}
\end{figure*}

Below we analize the dynamics and the off-center cooling process
triggered by a single AGN outburst. We focus here on the conical
outflows, the cylindrical ones giving qualitatively similar results.
This idealized case will be used to better understand the physical
nature of the non-linear density perturbations which lead to
widespread cooling.  We expect that in a more chaotic and realistic
ISM, perturbed by the intermittent action of AGN feedback, extended
cooling would be facilitated. This is the subject of Section
\ref{sec:feedback}.  In the following we focus on the gas that cools in the
``off-center region'' defined as: $z>500$ pc, $R>150$ pc.
We exclude the region near the symmetry $z$-axis because
the 2D cylindrical grid used in our simulations might lead in
spurious enhanced cooling along the axis. Indeed, the formation of cold
gas along the jet axis is a common result of our simulations. This is
physically reasonable because when the cavities generated by the AGN
jets move buoyantly toward larger distances, they trigger a vortex flow
around them, which compresses the hot gas toward the $z$-axis,
significantly lowering the local cooling time (see also
\citet{mb2008,revaz2008}; Brighenti, Mathews \&
Temi(2015)). Preliminary 3D calculations
verify the formation of the $z-$axis cold filament (C. Melioli, in
preparation), but we still prefer to be conservative in this work and focus
on the region far from the symmetry axis for most of our analysis.

We also neglect the gas which cools at the very center. The suppression of
the total cooling rate in galaxies and clusters relates to the
so-called cooling flow problem \citep{mcnamara2007} and will not be
discussed here. For calculations about
AGN heating in ellipticals see \citet[e.g.][]{gasp12},
which use a similar feedback scheme.

\subsection{AGN-cooling in the Central Galaxy}
\label{sec:central}

\begin{figure*}
\begin{center}
\adjincludegraphics[trim=1.8cm 2.4cm 1.2cm 1.cm, clip, max width=1.05\textwidth]{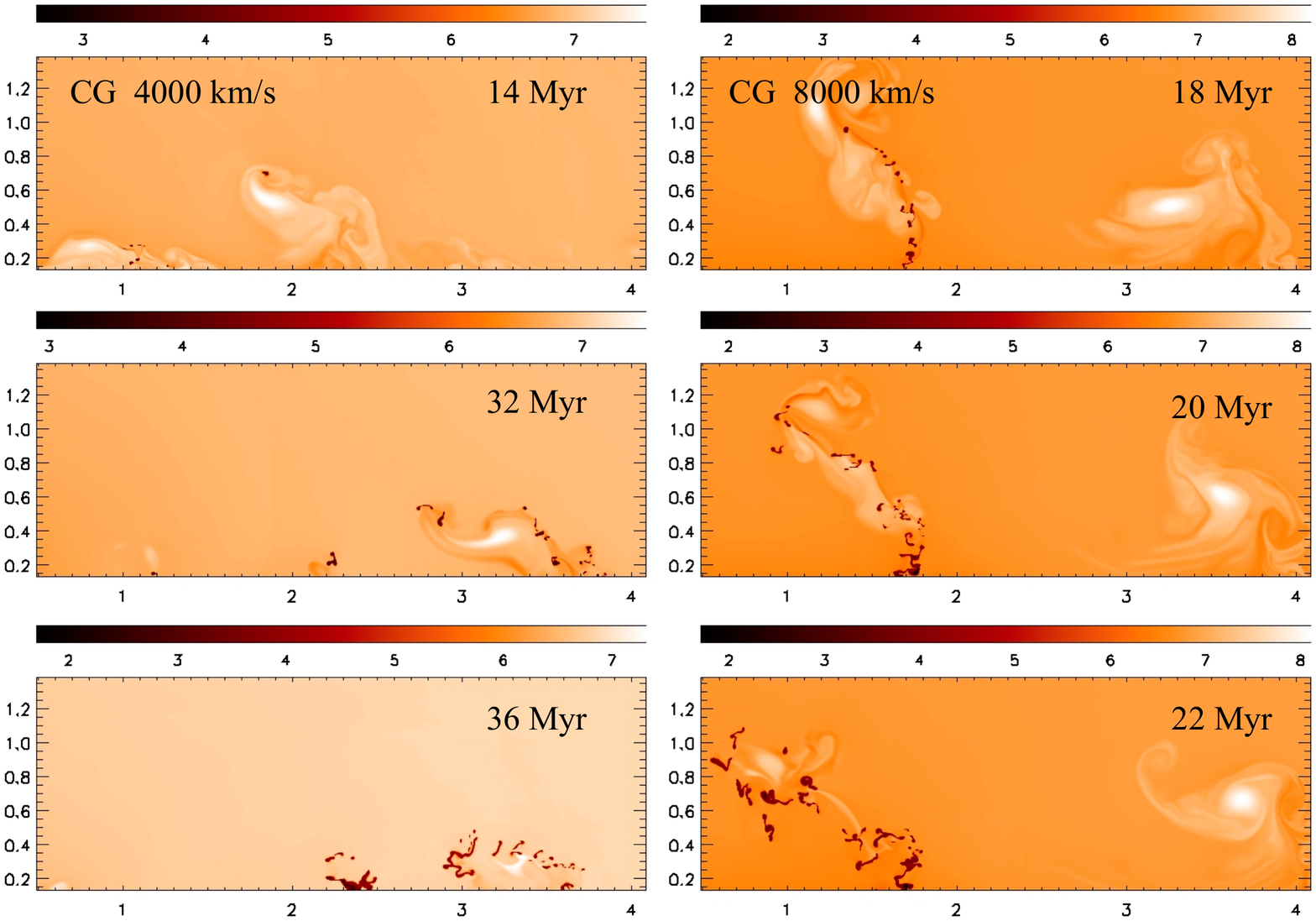} 
\end{center}
\caption{{\it Left column:} temperature map for model {\it CG4000} at three
 different times: $t=14,\, 32,\, 36$ Myr, from top to bottom. The {\it z-}axis is horizontal,
 the {\it R-}axis is vertical. Units are kpc. Time is
 measured since the switch on of the AGN outflow (which lasts 2
  Myr). {\it Right
    column:} same for model {\it CG8000}. The temperature is shown at 
$t=18,\, 20,\, 22$ Myr, to follow a strong cooling episode.
Notice that the color scale differs for every panel.
}
\label{fig:tCGiso}
\end{figure*}

In the top panel, central column of Figure \ref{fig:mcold_con} we show
the time evolution of the cold gas mass located off-center
($z>500$ pc, $R>150$ pc), for the single conical outflow
models. Times are measured since the activation of the outflow. 
For the {\it CG}, the three models with $v_{\rm jet}\,=\, 2000,\;
4000,\, 8000$ km s$^{-1}$ are able to cool a substantial amount of gas,
$10^4 - 10^7$ M$_\odot$,
at large distances from the center. A weak trend with $v_{\rm jet}$ is
evident from Figure \ref{fig:mcold_con}: the larger is $v_{\rm jet}$, the
larger is the mass cooled off-center. As already found in \citet{bm02}, the
AGN feedback generates non-linear, almost isobaric perturbations in
regions of sustained compression, which trigger intermittent and
widespread cooling throughout the region with $t_{\rm cool}/t_{\rm ff}\la
10$. Because of the large overdensities formed in converging flows, it
seems improper to describe this cooling process as a thermal instability.
This agrees with recent studies of cooling in galaxy clusters by
\citet{McCourt12,gaspari2012b}. Contrary to these latter works, however, we do
not include an artificial source of turbulence nor a distributed
source of heating. Generally, the region over which cooling
occurs is more extended when the power of the outflow is larger.

Interestingly, cooling can occur long after the AGN outflow ceases
(see also Brighenti, Mathews \& Temi, 2015).
For instance, for models {\it CG4000} and {\it CG8000} off-center
cooling episodes happen at $t\sim 35$ Myr and $t\sim 50$ Myr,
respectively, while the outflow terminated at $t=2$ Myr. 
At these times, usually no sign of the AGN feedback (cavities,
shocks) would be detected with X-ray observations. Nevertheless,
subsonic motion with velocity up to $\sim 1/2$ of the sound speed is
still present in localized regions, especially in form of relatively
large eddies with typical size $\sim 0.5$ kpc. We note that these
features are well resolved in our simulations with $\sim 50-100$ grid
points. When exceptionally large eddies are present in the inner region of the
galaxy, they appear as weak X-ray cavities. This happens
infrequently in our simulations, with large outflow velocities and
earlier times making this event more likely.

The spatial distribution of the cold gas in the off-center region at
several times, for models
{\it CG4000} and {\it CG8000}, is displayed in Figure \ref{fig:tCGiso}. The cold
blobs generally arise in a region close to the outflow symmetry axis
($R\la 1$ kpc) and for $z\la 10$ kpc. 
The cold
gas is arranged in a large number of small clouds (actually, toroidal
structures, given the imposed 2D cylindrical symmetry), generally few 10s pc
in size. As expected \citep[see][]{koyama2004} 
this is close to our numerical resolution ($\Delta R = \Delta
z = 5$ pc), with the unfortunate implication that the evolution of the
cold clouds may be subject to serious numerical errors. We address some
of the numerical problems, such as overcooling, in the Appendix.
The off-center cooling process starts in a very localized fashion,
usually in just one or few zones (see, for example, the top-left panel
in Figure \ref{fig:tCGiso}). This is true for every resolution adopted
in our numerical experiments. 
Calculations using coarse resolution might be grossly in error.

Using a tracer passively advected with the flow, we are able to track down
the original location of the cooling gas. We find that most of the
cold gas found off-center comes from the inner region ($r\la 1$ kpc),
where the entropy is lower. This confirms the results by
\citet{li2014}.
Because the ISM (iron) abundance peaks in
the center, it also follows that the metallicity of the
cold gas is somewhat higher than the one of the surrounding hot gas.

In Figure \ref{fig:mcold_con} we also show, in the central row, the evolution of the cold
gas mass in the region $r=\sqrt{R^2+z^2}>500$ pc (that is, we include here the cold
filament along the $z-$axis) and the total cold gas
(bottom panels). 
The mass in the filament forming in the wake of the
outflow lies in the range $M_{\rm fil} \sim 10^4 - 10^6$
M$_\odot$\footnote{Note that here and in the following we consider only  the $z>0$ volume.},
although these values must be confirmed by high resolution 3D
simulations. The time evolution of the
total cold gas mass shows that the single AGN outburst is unable to
significantly suppress the cooling rate with respect to the pure
coling flow scenario. Only for the {\it CG8000} model the AGN outflow
temporarily
reduces the cooling rate by a factor of $\sim 2.5$ for a time $\sim
15$ Myr, after which the system recovers the standard cooling rate,
$\dot M \sim 10$ M$_\odot$ yr$^{-1}$.

\subsubsection{Cylindrical outflows}

We have also calculated several models with cylindrical outflows, in
which the velocity is directed along the $z-$axis. The time evolution
of the cold gas mass for the models with jet velocity 2000, 4000 and
8000 km s$^{-1}$ is shown in the first column of Figure
\ref{fig:mcold_con}. The results are qualitatively similar to the
conical outflows, with a tendency to generate more off-center cold
gas. The reason is that cylindrical outflows are less efficient to
distribute their kinetic energy in the inner region of the galaxy,
therefore a larger volume of the system hosts an almost unperturbed
cooling flow. Also, it is well known that conical AGN outflows get
collimated by the hot gas pressure as they propagate outward, becoming
effectively cylindrical at some distance from the center
\citep[cfr.][]{brighenti2006}. 
Given the similarity of models with conical and cylindrical outflows
(and the uncertainties about real AGN outflows), in the following we
limit the discussion to conical jets.

\begin{figure}
\begin{center}
\adjincludegraphics[trim=8.2cm 1.8cm 0.8cm 0.7cm, clip, max width=0.81\textwidth]{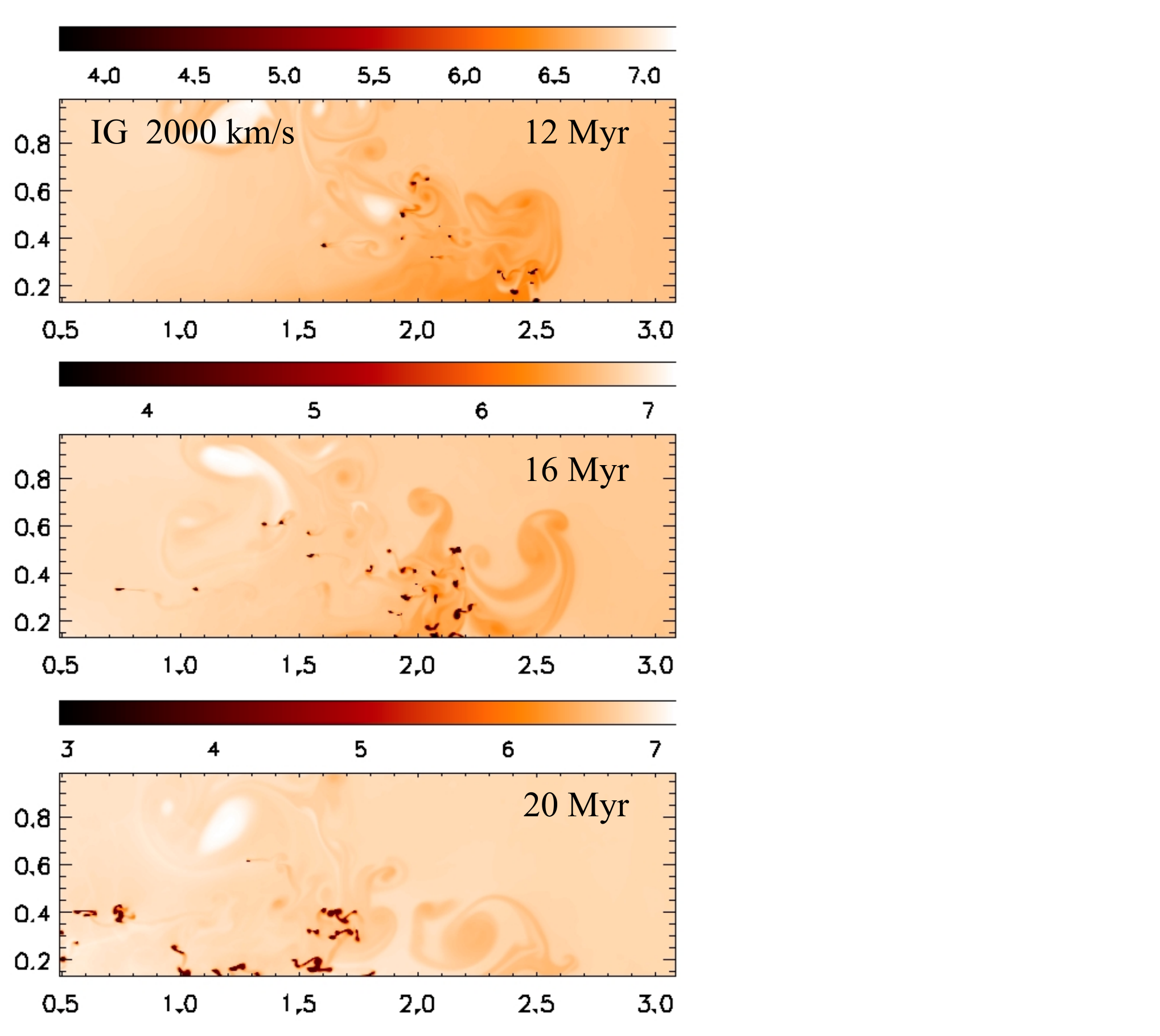}  
\end{center}
\caption{Temperature maps for the isolated galaxy ``single event'' and $v_{\rm jet} =
  2000$ km s$^{-1}$. Top panel: $t=12$ Myr; middle panel: $t=16$ Myr;
  bottom panel: $t=20$ Myr. The {\it z-}axis is horizontal,
 the {\it R-}axis is vertical. Units are kpc. Notice that the color scale differs for every panel.
}
\label{fig:tiso}
\end{figure}

\subsection{AGN cooling in the Isolated Galaxy}
\label{sec:isolated}

As pointed out in Section \ref{subsec:timescales}, 
the fundamental difference between
the {\it IG} and the {\it CG} models is that in the former the whole hot ISM is
generated by the stellar mass loss of the galactic (old) stellar
population. Therefore, the mass and the average density of the hot
gas is much lower than in the {\it CG} case. As a result, the ratio $t_{\rm
  cool}/t_{\rm acc}$ for the {\it IG} is everywhere larger by a factor of $\ga 10$
than for the {\it CG} (see Figure \ref{fig:ratios}). According to
the criterion by \citet[][]{Sharma12} and \citet{McCourt12} we should not
expect spatially extended cooling for this model.
This conjecture does not pass the test of our simulations.
In fact the model {\it IG2000} gives
rise a substantial mass of cold gas at $t\sim 5-25$ Myr (top-right
panel of Figure \ref{fig:mcold_con}). We will see below that in case
of repeated AGN outbursts, even moderate outflows with velocity of
$1000$ km s$^{-1}$ trigger significant off-center cooling.
The spatial map of the cold gas for {\it IG2000} is shown in Figure
\ref{fig:tiso}. The off-center cooling process initiates at $t \sim 5$
Myr, with small cold blobs forming at $(R,z)\sim (0.2, 2)$ kpc and
$(R,z)\sim (0.4, 2.1)$ kpc, where the time ratio $t_{\rm  cool}/t_{\rm
  acc} \sim 60$, just before the AGN activation\footnote{Just before the time
  of the first off-center cooling episode, the ratio $t_{\rm
    cool}/t_{\rm acc}$ near the cooling region has values in the range
  [20-50]. This is due to the advection of low entropy material from
  the center by the AGN outflow.}. As time goes by, more
gas cools and moves away from the original formation location. At later
times, as shown in Figure \ref{fig:tiso}, a multitude of cold clumps
can be found for $(R,z)\la (0.7, 2.5)$ kpc. As for the {\it CG}
simulation, the emergence of the off-center cooling depends on two
effects. First, the outflow transports (low entropy) gas initially at the center to larger
radii \citep[see also][]{li2014}. Second, the dynamical evolution of
the AGN outflow generates regions of sustained compression (where
$\nabla\cdot v < 0$), which gives rise to high contrast density inhomogeneities. 

As in the model {\it CG}, a massive cold filament is created along the axis
of the jet, with mass $\sim 10^4$ M$_\odot$, for models {\it IG2000} and
{\it IG4000}. No gas cools instead for $r>500$ pc in the case of the more powerful
model {\it IG8000} (middle-right panel of Figure \ref{fig:mcold_con}).
Evidently, the AGN heating is strong enough to completely suppress
off-center cooling, at least for the timespan probed by our simulation.

We notice that the average total cooling rate in the 50 Myr following
the AGN outflow, for models {\it IG1000, IG2000, IG4000} and {\it IG8000}, is
$0.5, 0.45, 0.32, 0.24$ M$_\odot$ yr$^{-1}$ (see bottom-right panel of
Figure \ref{fig:mcold_con} - model {\it IG1000} is not shown).  These values
reflect the (temporarily) suppression of cooling due to AGN activity,
being the cooling rate of the pure cooling flow model $\dot M_{\rm
  cool} \sim 0.8$ M$_\odot$ yr$^{-1}$ .

\subsection{AGN cooling in the Low Mass Galaxy}
\label{sec:lowmass}

We describe only very briefly the single outflow models for the low
mass galaxy. Because of the lower hot ISM density, widespread cooling is less efficient for this
system. However, we shall see below (Section \ref{sec:lowfeed}) that when repeated
AGN events are considerated, even the {\it LM} model undergoes significant off-center
cooling.

For the single outflow experiments, we find 
that only {\it LM2000} shows extended cooling, for $t\sim 10-25$
Myr, when $\sim 10^3 - 3\times 10^3$ M$_\odot$ are present in the
off-center region. Few small blobs develop at $z\sim 4, \; R\sim 0.5$
kpc. As usual, a cold filament also forms along the
$z-$axis, with mass $\sim 3 \times 10^3$ M$_\odot$.

\section{Spatially Extended Cooling: AGN feedback}
\label{sec:feedback}

The models presented in Section \ref{sec:single} showed that the
dynamical interaction of a single AGN-driven outflow with a smooth and
regular ISM generates density (entropy) perturbations which lead to
(sometime recurrent) widespread gas cooling. Understanding these
idealized calculations has been preparatory to the study of the more
realistic problem of repeated AGN outbursts, the subject of this Section.

The intuitive expectation is that in an ISM continuosly stirred by the
AGN feedback activity the generation of non-linear
perturbations is facilitated, with the result of more intense
cooling episodes. This is indeed what the simulations presented below show.

\subsection{Feedback in Central Galaxies}
\label{sec:cgfeed}

\begin{figure*}
\begin{center}
\adjincludegraphics[trim=0.7cm 9.6cm 1.5cm 8.3cm, clip, max width=1.\textwidth]{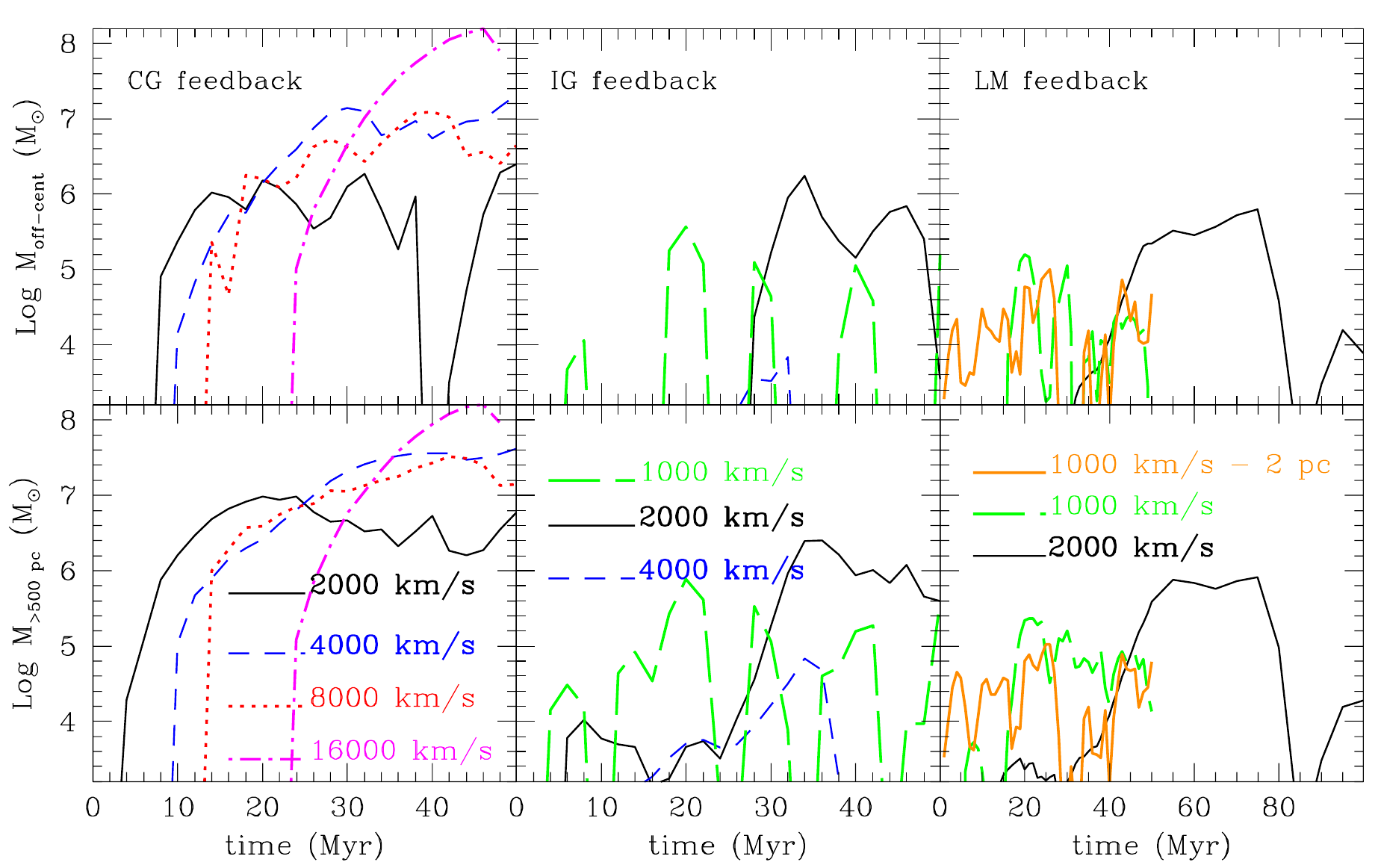} 
\end{center}
\caption{{\it Upper panel:} time evolution of the cold gas mass
  in the off-center region ($R>150$ pc, $z>500$ pc) for feedback models
  {\it CG-FB2000}, {\it CG-FB4000},  {\it CG-FB8000}, {\it CG-FB16000}
  (left panel), {\it IG-FB1000}, {\it IG-FB2000}, {\it IG-FB4000}
  (middle panel) and for {\it LM-FB1000} and {\it LM-FB2000} (right panel,
  notice that the time axis extends to 100
  Myr in this latter case). {\it
    Lower panel:} as above for the total cold gas mass in the
    off-center region $r>500$ pc.  
  }
\label{fig:mcold_feed}
\end{figure*}

\begin{figure*}
\begin{center}
\adjincludegraphics[trim=1.8cm 3.cm 2.3cm 2.5cm, clip, max width=1.\textwidth]{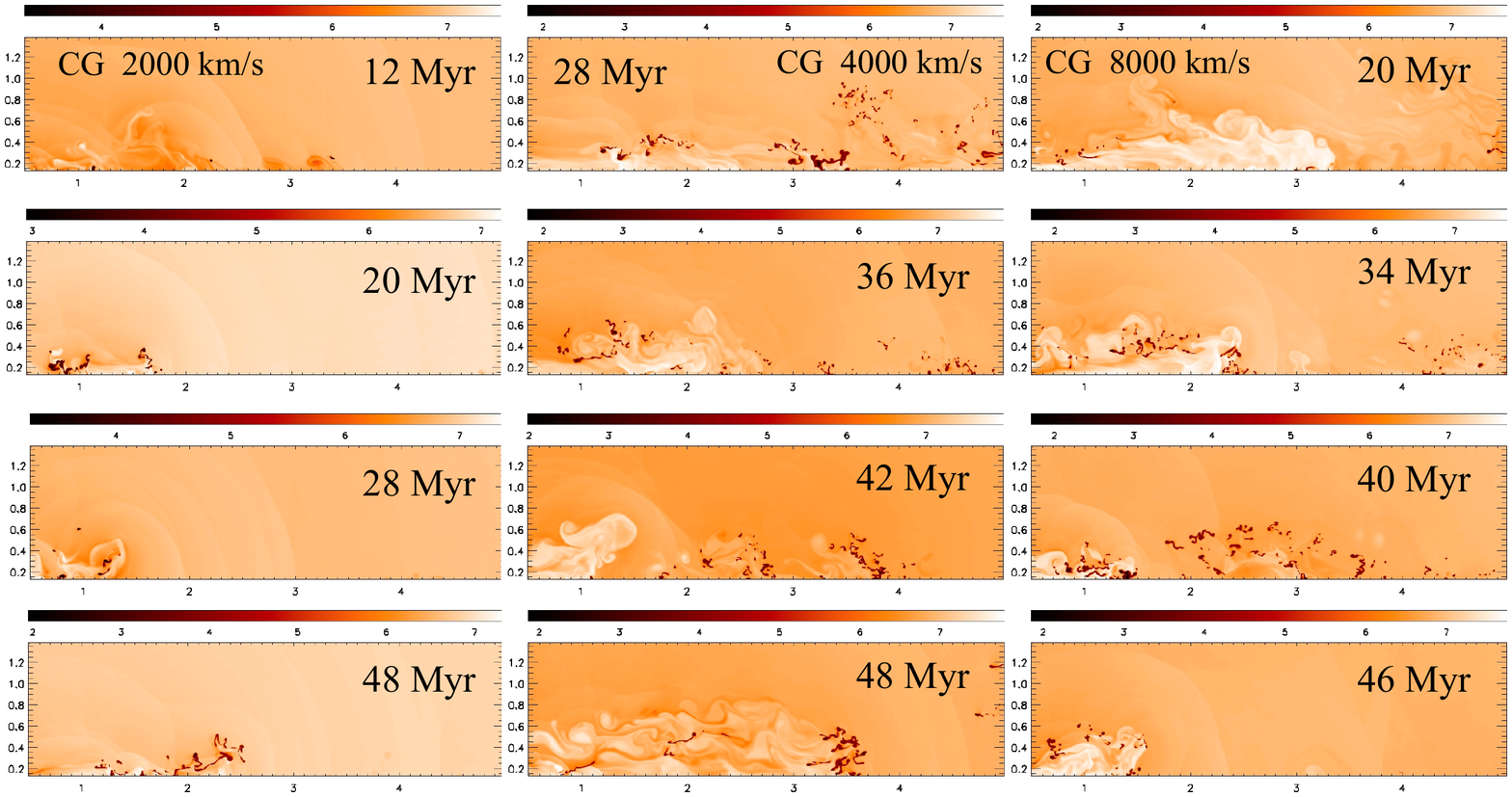} 
\end{center}
\caption{{\it Left column:} temperature maps for model {\it CG-FB2000} at four
  different times: $t=12,\, 20,\, 28,\, 48$ Myr, from top to bottom. The {\it z-}axis is horizontal,
 the {\it R-}axis is vertical. Units are kpc. Time is
  measured since the switch on of the AGN outflow (which lasts 2
  Myr). {\it Central column:} temperature maps for model {\it CG-FB4000},
  shown at 28, 36, 42 and 48 Myr. {\it Right
    column:} temperature maps for model {\it CG-FB8000} at $t=20,\, 34,\, 40$
  and $46$ Myr. 
Notice that the color scale differs for every panel.
}
\label{fig:CGfeed}
\end{figure*}

In Figure \ref{fig:CGfeed} we show the off-center temperature maps
for the
feedback models of the {\it CG} galaxy, with outflow velocities of 2000,
4000 and 8000 km s$^{-1}$. The dark spots in every panel indicate gas
cooled to $T<10^4$ K. In Figure \ref{fig:mcold_feed} the mass of cold
gas in the various regions, as defined
in the previous sections, is shown for most models.
The first result revealed by these images is that a persistent cold
ISM phase is naturally present, regardless on the feedback
details. 
Of particular interest is model {\it CG-FB16000} --- with the fastest
outflows among our simulations --- shown in Figure
\ref{fig:mcold_feed} with the magenta dot-dashed line. In this system
the AGN heating is strong enough to suppress almost completely the cooling at
the center of the galaxy (where all the cooling in the classic cooling
flow picture would occur). Despite that, a strong off-center cooling
episode happens at $t\sim 24$ Myr and generates a total cold
gas mass $M_{\it blob}\sim 1.5\times 10^8$ M$_\odot$. 

The network of blobs and filaments occupies a region
typically few kpc in size and has mass $M_{\rm blob} \sim 10^6 - 10^7$
M$_\odot$. We are not in the position to calculate the fraction of
this gas in molecular, neutral or photoionized state. It is likely, however,
that all these components are simultaneously present (see our Discussion
below), as supported by observations \citep[e.g.][]{werner2014,david2014}.
The spatial distribution of the cold gas generally
correlates with the direction of the outflows and with the
disturbances induced in the hot gas. It is worth noting that the
brightness of warm gas clumps is expected to decrease with the
distance from the center, in pace with the decrease of the pressure.
The aspect and amount of the AGN-triggered cold
phase are in broad agreement with warm/cold gas observations in massive
ellipticals/galaxy groups \citep{caon2000, werner2014}. We believe
that this is the most robust and relevant result of our simulations.
The velocity structure of the spatially extended cold phase is
discussed in Section \ref{sec:vdisp}.

The filament of cold gas forming along the $z-$axis is always present
in the {\it CG} simulations, with mass which varies between few $10^5$
to few $10^7$ M$_\odot$. 
The filament has often a clumpy appearance,
with density variations within it of $2$ orders of magnitude. The
velocity of the cold gas is generally negative (especially at late
times) with typical values of $100-200$ km s$^{-1}$, 
but outflowing portions of the filament are also present. 

The physics of a multiphase ISM and the interaction between the
phases is very complex, a process poorly understood even in our own
Galaxy \citep[e.g.][]{cox2005}, and we have no chance to give a
consistent description here. As for the single outflow case, 
the cold material in our feedback simulations
condenses in very small structures, usually few numerical zones in
size. Resolution tests indicate that this is true even when the cell
size is reduced by a factor of 5 (that is, adopting $\Delta R = \Delta
z = 1$ pc). The volume filling factor in the
central region where cold gas is found is always very small, usually
less than 2-3\%. 
Off-center cold gas usually spans temperatures in the range $10^3-10^4$
K, with only a few percent undergoing cooling below $100$ K.

Initially, the cold blobs share the same turbulent velocity field of the hot gas
out of which they form. Then, they start to fall toward the center of
the system, which is reached in about a dynamical time ($\sim 10^7$
yr); this is the typical lifetime of an off-center cloud or filament.

\subsection{Feedback in the Isolated Galaxy}
\label{sec:igfeed}

\begin{figure*}
\begin{center}
\adjincludegraphics[trim=1.7cm 5.2cm 2.2cm 2.8cm, clip, max width=1.\textwidth]{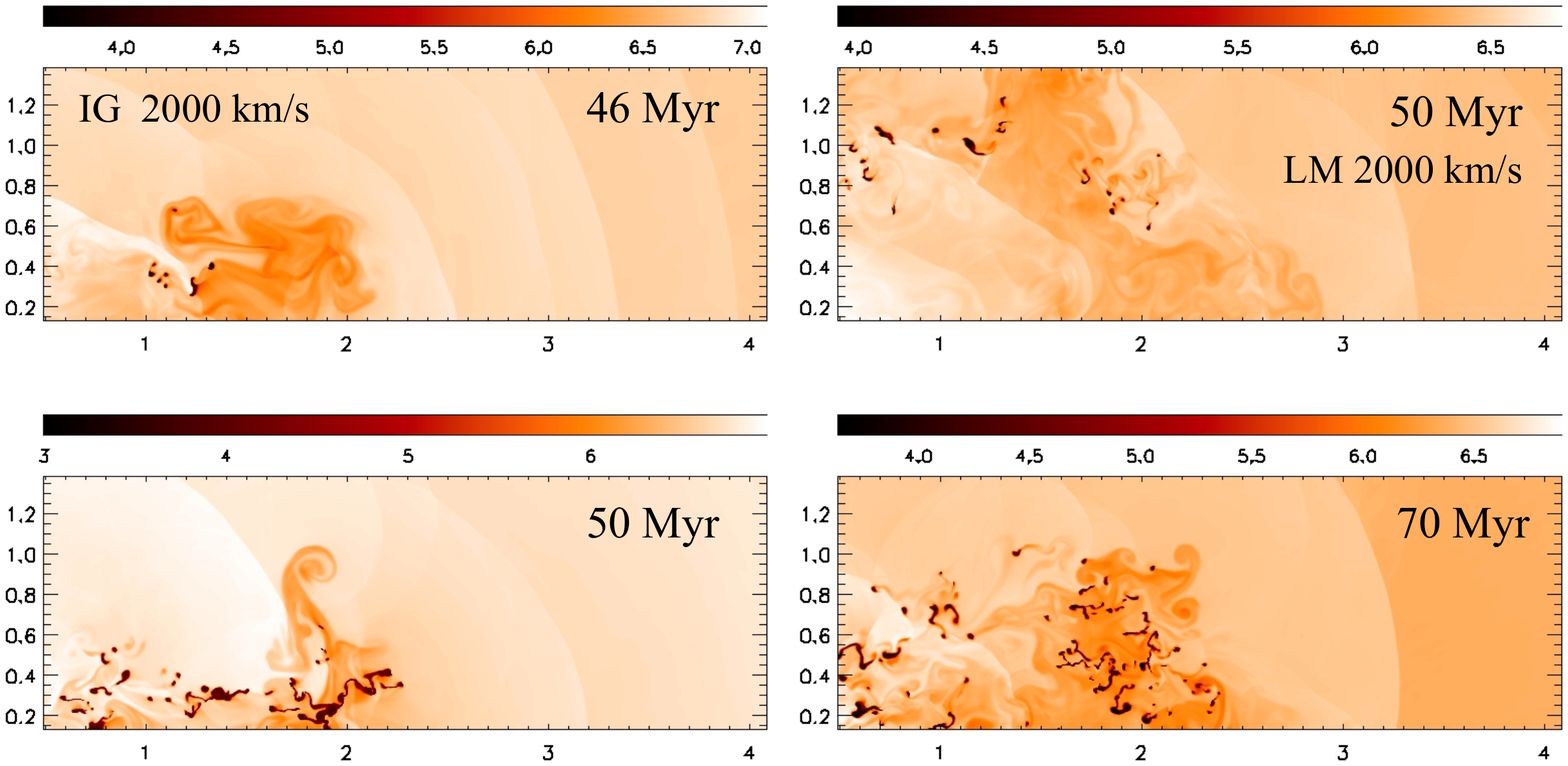} 
\end{center}
\caption{{\it Left column:} temperature maps for {\it IG-FB2000} model
  of the isolated elliptical. The {\it z-}axis is horizontal,
 the {\it R-}axis is vertical. Units are kpc.  {\it Right
    column:} same for {\it LM-FB2000} model. Notice that the color scale differs for every panel.
}
\label{fig:IGLMfeed}
\end{figure*}

The isolated galaxy, although more optically luminous than the {\it CG}, has a
much lower amount of hot ISM. As discussed earlier on, this leads to 
a longer $t_{\rm cool}$ and a larger $t_{\rm cool}/t_{\rm
  acc}$. Correspondingly, the entropy is higher. 
However, this only slightly hinders the spatially cooling
process (see also Section \ref{sec:isolated}). We find that both models
{\it IG-FB1000} and {\it IG-FB2000} generate an extended cold ISM
phase, whose physical properties vary in time (see Figures
\ref{fig:mcold_feed} and \ref{fig:IGLMfeed}). Total masses of cold gas $M_{\rm cold}\sim
10^5-10^6$ M$_\odot$ are common in the off-center region ($R>500$ pc,
$z>150$ pc). For model {\it IG-FB4000} only little gas cools in this
volume, $M_{\rm cold}\sim 3.8 \times 10^3$ M$_\odot$, and there is
only one episode of extended cooling at $t\sim 30$ Myr.
Of course it is possible, perhaps likely, that more gas would condense
at times $>50$ Myr. We did not calculate a feedback model for the {\it
  IG} with $v_{\rm jet} = 8000$ km s$^{-1}$, but it seems plausible that recurrent
outflows with such a velocity prevent off-center cooling completely,
as the single outflow does for the {\it IG8000} model (see Section
\ref{sec:isolated}).

The cold gas mass in the $z-$axis filament is comparable with that in the
off-center blobs, varying between zero and $\sim 7\times 10^5$
M$_\odot$ in models {\it IG-FB1000} and {\it IG-FB2000}, while for
{\it IG-FB4000}  it reaches $\sim 7\times 10^4$ M$_\odot$ at its
maximum (cfr. top and bottom panels in Figure \ref{fig:mcold_feed}).

\subsection{Feedback in the Low Mass Galaxy}
\label{sec:lowfeed}

Results for this model, which has an optical luminosity close to the
$L^\ast$  parameter of the luminosity function, are relevant for a
large number of normal elliptical galaxies. 
Before the onset of the AGN, the (hot) ISM mass within the effective radius is $\sim 1.2
\times 10^7$ M$_\odot$; within $r\sim 2$ kpc the ISM is slowly inflowing as
in the classical cooling flow theory, while at larger radii the SNIa
heating generates a highly subsonic outflow, with average velocity 
$v \sim 10$ km s$^{-1}$.
Is interesting that while gas cooling occurs also where $t_{\rm
  cool}/t_{\rm acc} \gg 10$, it is however limited to the ISM inflowing region
($r\la 2$ kpc, Figure \ref{fig:IGLMfeed}), within which the initial ISM mass
is $\sim 6.9 \times 10^6$ M$_\odot$. We also calculated few models with
a higher SNIa rate (and therefore a smaller or no inflowing region,
although the outflow velocity is still very subsonic) to test
the reasonable hypothesis that global inflow (except for the outflow
region, which occupy a small volume of the system) is a necessary condition for
widespread cooling. Clearly, non linear density perturbations moving 
outward, toward lower pressure regions, would expand, lengthening
$t_{\rm cool}$ and delaying or frustrating the cooling process. In
these ``windy'' galaxies we find that no gas cools\footnote{In these
  systems hot gas could in principle still cool through the dust assisted cooling
  process described in \citet{mb03}.}.

In Figure \ref{fig:mcold_feed}, right column, the time
evolution of cold gas mass in the off-center region and in the
$r>500$ pc region is plotted. For models {\it LM-FB1000} and {\it LM-FB2000}
an extended cold ISM phase naturally forms within $r\la 1$ ($r\la 2$)
kpc for {\it LM-FB1000} ({\it LM-FB2000}), again in an elongated
region around the $z-$axis. For the high outflow velocity model {\it
  LM-FB4000} no gas is able to cool. In this case the AGN heating
offsets radiative cooling for the (short) time covered by our
simulation (50 Myr). It is entirely plausible that some cooling would
happen at later times \citep[see][]{gasp12}, but we did not explore
this possibility here. Cold gas masses of $1-5 \times 10^5$ M$_\odot$
are typical, with periods of much lower or no cold gas found in the
galaxy.

As a numerical convergence test we run a model identical to {\it
  LM-FB1000} but with a grid resolution of 2 pc. The results are
illustrated in the right panels of Figure \ref{fig:mcold_feed} by the
orange lines. While the detailed pattern of the cold gas mass evolution
differs, the important result --- the generation of an extended, $\sim
10^5$ M$_\odot$ cold ISM phase, even where $t_{\rm cool}/t_{\rm acc} >
10$ --- is validated.

\section{Dust in the hot and cold gas}

\subsection{Central Galaxy}
\label{sec:centraldust}

As mentioned in the Introduction, the dust content of the cold gas is a
crucial test for its origin. 
The necessary first step is to understand the grain
evolution in the hot ISM. This has been investigated in detail by
\citet{tsai1995,tsai1996}, \citet{temi2003}, \citet{dwek1990},
\citet{donahue1993}, \citet{hensley2014}. Stellar ejecta continuosly supply dust
to the ISM, at a rate of $\delta_\ast \dot M_\ast \approx 10^{-3}
(M_\ast/10^{11}\; \rm M_\odot)$, assuming $\delta_\ast = 1/150$.
In \citet{tsai1995} and \citet{temi2003} it is
argued that stellar ejecta are quickly heated to the hot ISM
temperature \citep[see also][]{mathews1990} in a timescale $\la 10^6$ yr \citep[a fraction $\la 20$\% 
of it may remain cold, according to the calculations
by][]{parriott2008}. Dust grains, as they come in contact with $\sim 10^7$
K gas, are sputtered away with a timescale $\tau_{\rm sputt} \sim 0.03/n_p$
Myr (which translates in $\tau_{\rm sputt}\la 10^5$ yr in the galaxy core and $\tau_{\rm
  sputt}\sim 3 \times 10^6$ yr at $r=10$ kpc), 
short enough to make grains be disrupted essentially where they
originate. 

Our model for the dust evolution in the hot gas agrees with
the aforementioned calculations very well. Dust source and sink terms
result, for the {\it CG} model, in an
almost constant ISM dust-to-gas ratio $\delta \sim 3\times 10^{-6}$.
Before the outflow starts the
total mass of dust present in the gas in the whole system is $M_{\rm
  dust}\sim 7.4 \times 10^4$ M$_\odot$, of which only $\sim 5 \times
10^3$ M$_\odot$ reside in the central region within $r_{\rm e} = 10$
kpc. These numbers are very similar to those of the steady state model
for interstellar dust in NGC 5044, presented in
\citet[]{temi2007}, where it has been showed that the observed fluxes
at 70 $\mu$m and 170 $\mu$m are a factor $50-100$ higher than expected
if the only source of cold dust were the
stellar winds.
The excess dust in NGC 5044 is spatially
correlated to the warm ($T\sim 10^4$ K) gas
\citep[see also][]{gastaldello2009}.

Gas cooling out of the hot phase initially shares the aforementioned low dust
content. However, dust can grow in cold gas by accretion
of condensable elements onto preexisting grains, as outlined
in Section \ref{sec:models}. We find that dust accretion is never
significant for the cold blobs found in our simualtions. The increase
in the dust-to-gas ratio of cold gas is modest, leading to an average
$\delta_{\rm cold} \la 10^{-5}$ after 50 Myr, close to the dust-to-gas ratio
of the hot phase. We see little variation in the dust-to-cold gas
ratio between models.
The inefficient dust growth originates from both the relative low
density of the cooled gas clouds (usually $\la 50$ cm$^{-3}$)
and their short lifetime. Off-center cold blobs, in fact, are young
objects, with life span $\sim t_{\rm 
  dyn} \la 10^7$ yr, after which they settle in the galaxy core or in
the cold filament along the $z-$axis\footnote{We
  neglect here the potential role of hot gas rotation, which would
  generate a large disk of cold gas, several kpc in radius
  \citep[e.g.][]{brighenti1997}.}.

Therefore, a robust prediction of our calculations
is that internally produced, spatially extended cold gas is
very dust-poor.
Evidently, any spatially extended (and therefore
recently formed) dusty cold gas in isolated elliptical galaxies must
have a different origin, as for example cooling of dusty stellar
ejecta mixed with local hot gas \citep{mb03,temi2007} or accretion
of external gas \citep[e.g.][]{davis2011}. 
On the contrary, cold gas can reside in the $z-$axis filament long
enough, and with density large enough, 
to allow dust accretion be effective. At the end of the feedback
simulations, at $t=50$ Myr, the dust-to-gas ratio in the filament
reaches values of $\sim 5\times 10^{-3}$. For the same reason,
dust growth can also be an
important process in old cold gas located at the very center of a
large fraction of ellipticals, known to be dusty
\citep[e.g.][]{tran2001, martel2004, lauer2005, simoes2007,
  alatalo2013}.

We remark that the low dust content for the cooled hot gas might also
lead to a low $H_2$ and CO
content (with respect to the Galactic standards), even if the 
temperature dropped to very low values $T \la 100$ K \citep[see][]{cazaux2009}.
Furthermore, dustless cold gas would
not be effective in absorbing stellar radiation and would be hardly
detectable in dust extinction maps.
Some indication of dust-poor molecular gas in NGC 5044 is provided by
the ALMA observation presented in \citet{david2014}. Mainy molecular
associations have been found the inner $\sim 2.5$ kpc, for a total
molecular mass $\ga 5 \times 10^7$ M$_\odot$. The absorption map
presented in \citet{david2014} shows dusty patches that do not
correlate with the molecular structures. If the dust poor cooled gas is able to
form $H_2$ and $CO$ in $\la 10^7$ yr, molecular clouds in ETGs would
not provide much absorption or extintion, just like observed.

Only the cold
$z-$axis filament is expected to have a sizeable amount of dust but,
although the formation of this feature is entirely reasonable, we are
chary to study in detail its dust properties here, for the reasons explained above.

\subsection{Isolated Galaxy}
\label{sec:isolateddust}

The hot gas dust content in the standard CF model for the
IG is $2.7 \times 10^6$ M$_\odot$. 
This amount is more sizeable than in the corresponding
{\it CG} model because of the
lower average ISM density, which makes the grain sputtering process
less efficient ($\tau_{\rm sputt}\propto n^{-1}$), especially in the
outer galactic regions. Within $r=10$ kpc $M_{\rm dust, hot} = 1.4
\times 10^5$ M$_\odot$, while within $r_{\rm e}=6$ kpc we have
$M_{\rm dust, hot} = 3.6 \times 10^4$ M$_\odot$.
The dust-to-gas ratio in the hot ISM is now larger
than that of the {\it CG}, rising from $\delta_{\rm hot}\sim
10^{-5}$ in the very center, to $\delta \sim 10^{-3}$ at the effective
radius (6 kpc).

The higher $\delta_{\rm hot}$ leads to a higher
average dust-to-gas ratio for the off-center cold ISM phase,
$\delta_{\rm cold} = 3-10 \times 10^{-5}$, for both the single outflow
and the feedback models. Higher velocity outflows
typically generate higher $\delta_{\rm cold}$ because gas cools at
larger radii, where $\delta_{\rm hot}$ is also larger.
The situation is therefore similar to that depicted
for the {\it CG} model: spatially extended cold gas originated by cooling of the
hot phase is predicted to exist in real ellipticals with
AGN feedback. However, it is expected to be almost dustless; any
dust-rich cloud or filament has likely been accreted or ejected from
the nuclear region, where dusty cold gas is known to be present.

In the $z-$axis filament, the dust-to-gas ratio never exceeds $8\times
10^{-5}$. The lower value with respect to the {\it CG} model is mostly
due to the lower density in the {\it IG} filament, which makes the
grain growth process relatively inefficient.

\subsection{Low Mass Galaxy}
\label{sec:lowdust}

The {\it LM} galaxy has the lowest average hot ISM density among our
models. This makes the sputtering time longer and the dust content 
more substantial than in the {\it CG} and {\it IG} systems, with an average dust-to-gas ratio $3.8
\times 10^{-3}$ within 10 kpc and $1.4\times 10^{-3}$ within $r_{\rm
  e}$, at the time of the AGN activation. The dust mass in
the hot gas is $M_{\rm dust, hot}\,=\, 1.7\times 10^4 \, ($or $1.3\times 10^5)$
M$_\odot$ for $r\le \,3.16 \, ($or $10)$ kpc, respectively.
The cold clumps and filament generated by the AGN outbursts are both 
characterized by average $\delta_{\rm blob,fil} \sim 10^{-4} -
10^{-3}$, with the largest values found for model {\it LM-FB2000}.

\section{Gas velocity dispersion and cold inflow/outflow}

\subsection{Central Galaxy}
\label{sec:vdisp}

It is interesting to investigate the gas velocity dispersion generated
by the AGN outflow. The chaotic motion triggered by the outflow
expansion and the following buoyant motion of low-density regions
cause the density perturbations leading to widespread cooling. Also, it
promotes the circulation of the metals injected by the SNIa occurring
in the galaxy
\citep[e.g.][]{rebusco2006, gasp11a, gasp11b, gasp12}. Recent X-ray
observations provide estimates on the turbulent velocity in clusters
and groups, albeit with large uncertainties 
\citep[]{werner2009, sanders2011, dePlaa2012}. For NGC 5044
and NGC 5813 the latter authors find $320 < v_{\rm turb} < 720$ km
s$^{-1}$ and $140 < v_{\rm turb} < 540$ km s$^{-1}$ respectively,
measured in the central $\sim 10$ kpc region.

\begin{figure}
\begin{center}
\adjincludegraphics[trim=1.3cm 6.5cm 2.5cm 5.8cm, clip, max width=0.505\textwidth]{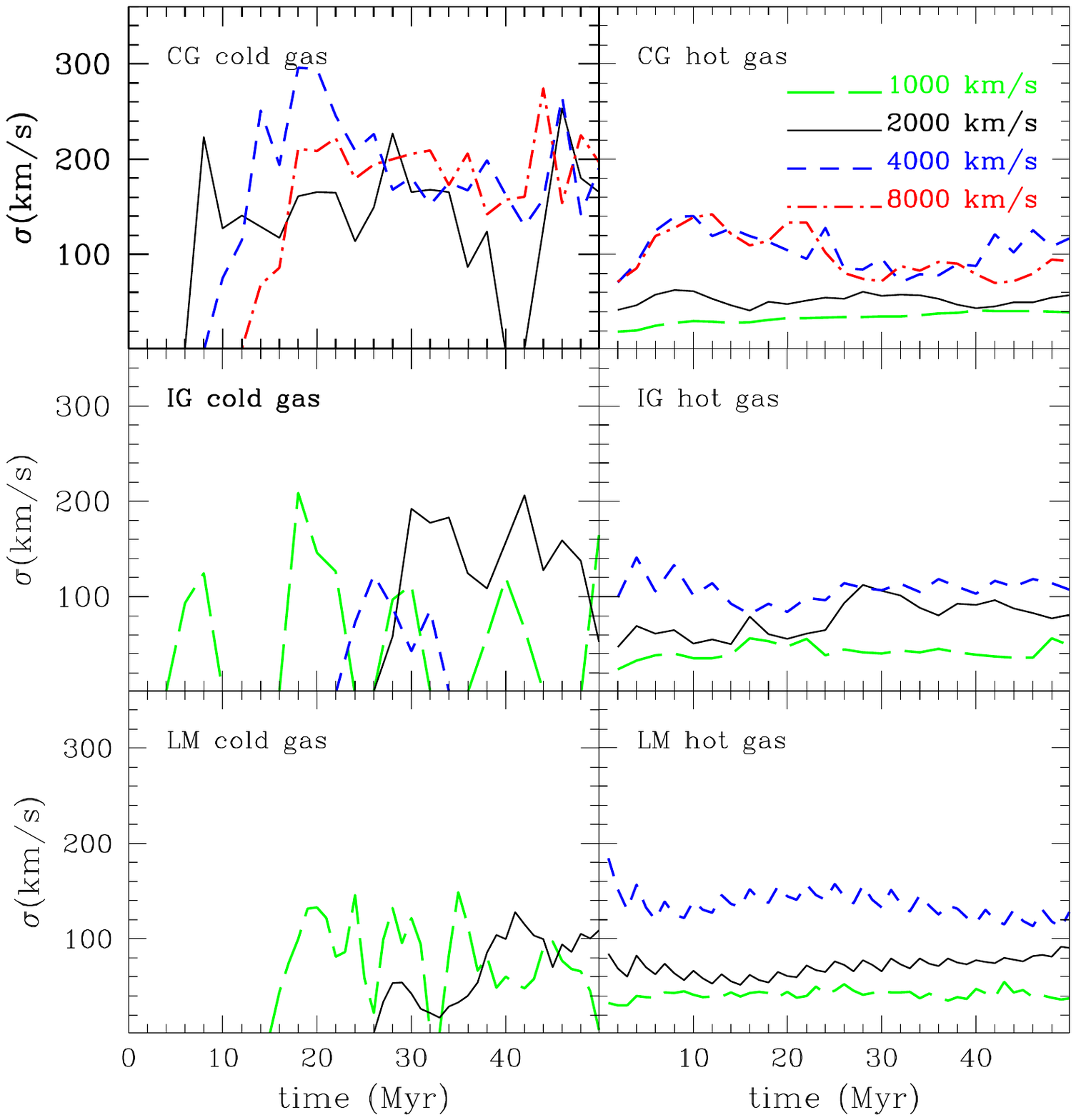}
\end{center}
\caption{{\it Left panels:} time evolution of cold gas velocity
  dispersion for gas in the off-center region defined by $0.5<z<5$ kpc,
  $0.15<R<3$ kpc. 
  The {\it CG-FB}, {\it IG-FB} and {\it LM-FB} models are displayed in the
  top, middle and bottom panels, respectively. For each system we show
  the feedback model with $v_{\rm jet} = 1000, 2000, 4000$ km s$^{-1}$
  with long-dashed green, solid black and dashed blue lines (for
  the CG model we also show the $v_{\rm jet} = 8000$ km s$^{-1}$ model
  with a dot-dashed red line),
  respectively.  {\it Rigt panels:} same for the hot gas.
}
\label{fig:vdisp}
\end{figure}

In the pure cooling flow model, before the onset of the AGN feedback,
the hot gas 3D, mass-weighted velocity dispersion is extremely low, $\sigma_{\rm
  hot}\sim 14$ km s$^{-1}$ (here and in the following, the
velocity dispersion is measured in the region $0.5<z<5$ kpc, $0.15<R<3$
kpc, corresponding to the off-center region where the numerical resolution is
higher)\footnote{We estimate the mass-averaged velocity dispersion as $\sigma^2 =
  \sigma_z^2 + 2\sigma_R^2$, where the $z-$component is $\sigma_z^2 =
  \int{(v_z-\hat{v_z})^2\rho dV}/\int{\rho dV}$ and the integrals are
  calculated on the volume defined in the text. The $R-$component is
  computed in a similar way.}. This value for $\sigma_{\rm hot}$
  essentially reflects the (smooth) radial velocity gradient typical
  for cooling flows in the region considered.
The single outflow event increases the hot gas turbulence, 
with typical velocities 
$\sigma_{\rm  hot}\sim 15, \, 35, \, 60, \, 100$ km s$^{-1}$ for
{\it CG1000, CG2000, CG4000} and {\it CG8000} respectively, 
measured few Myr after the AGN outburst. This chaotic motion
slowly decays, reaching values $\sigma_{\rm  hot}\sim 15, \, 30, \, 30, \, 40$ km s$^{-1}$
at $t=50$ Myr.
The cold gas, when present, usually shows larger velocities,
with typical values $\sigma_{\rm  cold} \sim 80-150$ km s$^{-1}$, 
essentially independent on
the outflow velocity. These numbers compare well with observed cold
gas velocity dispersion in massive ellipticals \citep[]{caon2000,werner2013}.

In Figure \ref{fig:vdisp} we show the velocity dispersion for the cold
and hot ISM for the feedback models. Again,
$\sigma_{\rm hot}$ generally increases with $v_{\rm jet}$
(although models {\it CG-FB4000} and {\it CG-FB8000} are almost
undistinguishable). Typical  $\sigma_{\rm hot}$ values range between
50 and 100 km s$^{-1}$. The obvious difference with respect to the
single outflow models is that $\sigma_{\rm hot}$ stays approximately
constant in time, as expected for a self-regulating feedback scenario.

We also find that $\sigma_{\rm hot}$ is strongly dependent on the
position. As expected, the turbulence is much stronger near the jet
axis. If we measure $\sigma_{\rm hot}$ in the sector $R\ge z$ (that is,
far from the jet axis) we get values a factor $1.5-2$ lower than in
the sector $R\le z$. This implies that any mixing and turbulent heating
should be much less effective in the region near the plane of symmetry $z=0$.

The values for $\sigma_{\rm hot}$ we find are significantly lower than those observed.
Although estimates for $\sigma_{\rm hot}$
in real galaxies are still sparse, this discrepancy may indicate that AGN
feedback is more powerful than the one considered in the present
paper \citep[see, e.g.][]{gasp12}. Stronger outbursts are indeed needed
to prevent the excessive total cooling rate observed in our simulations.
A further possibility to explain the observed large turbulent velocity
is the effect of merging and accretion into the system. For instance,
in NGC 5044 a pair of cold fronts are visible within the inner $\sim
70$ kpc region, likely generated by the accretion of a galaxy or a
small group \citep{gastaldello2009, david09}. The large ($\sim 150$ km
s$^{-1}$) peculiar velocity of the central galaxy with respect to the
mean group velocity supports this scenario.

Also for the feedback models we find that 
$\sigma_{\rm cold} \ga \sigma_{\rm hot}$. This is in part explained by
the tendency of the gas to cool in regions of localized stronger turbulence. In
addition, once cooled to low temperatures, the blobs decouple
dynamically from the hot gas (although their motion is never totally
ballistic\footnote{The fact that single blobs are not properly
  resolved in our simulations and the effect of numerical viscosity
  may play a role in maintaining a slight coupling between hot and
  cold gas. We did not investigate this in detail.}), and this results
in a further increase of $\sigma_{\rm cold}$. Overall, the recently
formed cold gas moves
in a somewhat chaotic way, with velocities similar to those found by
\citet{caon2000}. We note here that in our simulations no cold gas is
dragged to large radii from the central region. This indeed must be
expected, given the high density and small cross section of the cold
clouds, which make them almost immune to drag by hot gas or
entrainment. In models with large $v_{\rm jet}$ we do see cold blobs
moving outward together with the hot gas, but this is because the cold
gas forms from the outflowing hot gas and initially shares the same
motion. 

\begin{figure}
\begin{center}
\adjincludegraphics[trim=3.2cm 4.9cm 0.3cm 5.1cm, clip, max width=0.7\textwidth]{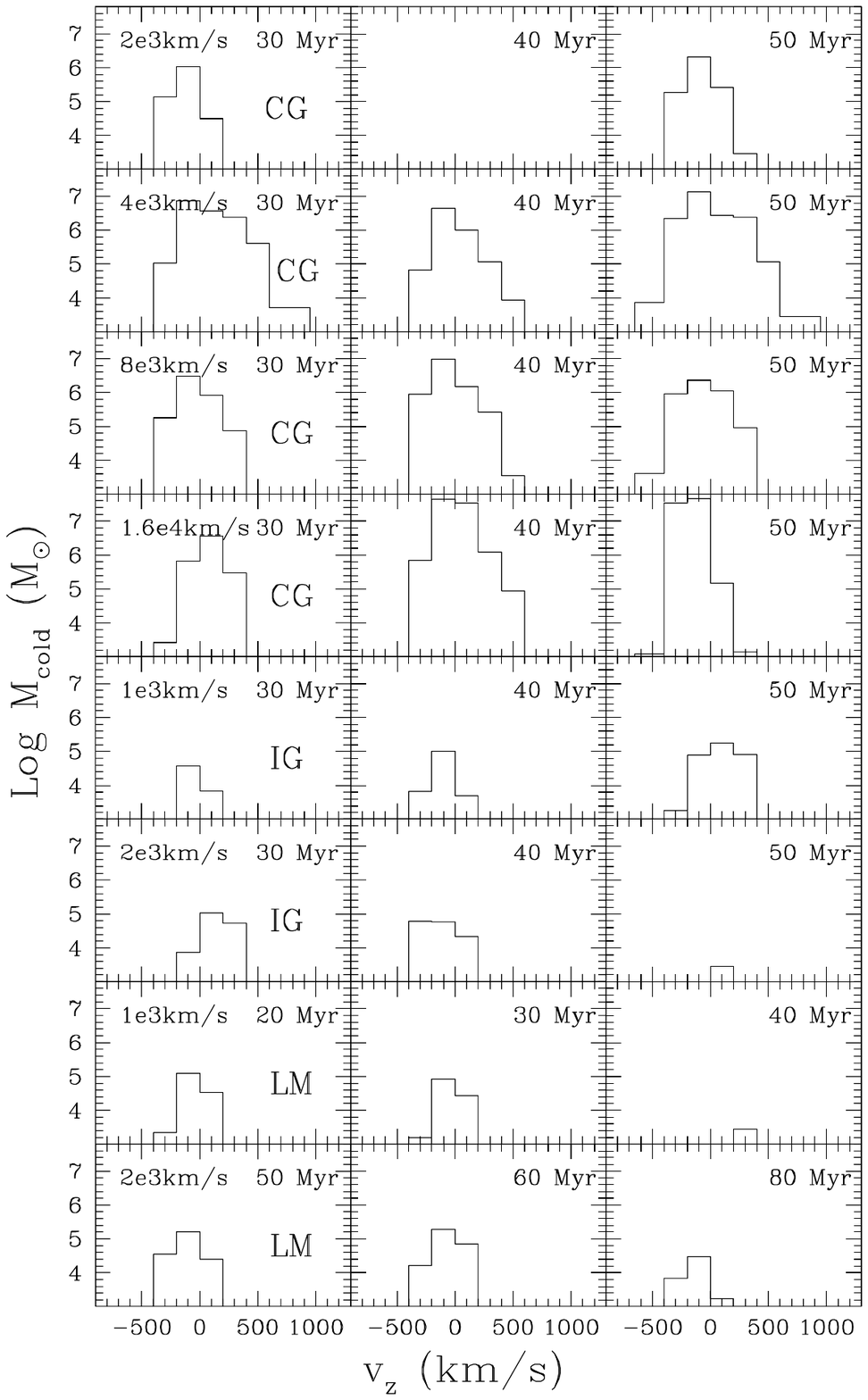}
\end{center}
\caption{Distribution of the cold gas in the z-axis velocity space for
  various feedback models at several times. Positive velocity indicate
  outflowing gas. Every row represents a model, indicated on the
  left panel of each row (for example in the top row is shown the {\it CG-FB2000} run).
}
\label{fig:histo}
\end{figure}

At any time cold gas clouds with both
positive (outflowing) or negative (inflowing) radial velocity are
present. Often
gas cools and condenses out from outflowing hot gas, resulting in a
cold outflow. We  note that molecular outflows are now commonly
detected in AGN
hosts \citep[e.g.][]{cicone2014,tadhunter2014}, albeit in systems different
than the group centered massive elliptical simulated here. 
In our scenario \citep[see also][]{li2014,Sharma12} the cold gas 
is not directly accelerated by the AGN,
but merely forms in the already forward moving hot gas\footnote{In
a very recent paper \citet{costa2014} have shown that hot gas
  cooling can generate cold outflows in QSO hosts.}. As for the mass
and spatial disposition, the distribution of the off-center
($0.5<z<5$ kpc, $0.15<R<3$ kpc) cold gas mass in the velocity space, shown
in Figure \ref{fig:histo} for several models, also varies in an
unpredictable way. For most models/times there is a slight tendency
for the cold gas to prefer outward motion, but the effect is
subtle. Typical velocities of $\sim$ few km s$^{-1}$, both positive
and negative, along the outflow symmetry axis are always present. It
is interesting that the cold gas velocity distribution does not depend
in an obvious way on the outflow velocity $v_{\rm jet}$ --- the model
{\it CG-FB16000} does not show a high velocity tail as might be
expected.

\subsection{Isolated and Low Mass Galaxies}

For these models, the hot gas velocity dispersion before the AGN onset
is $\sigma_{\rm hot}\sim 5$ km s$^{-1}$, somewhat less than in the 
{\it CG} case.

The single AGN outflow in the {\it IG} generates chaotic motion with
$\sigma_{\rm hot}\sim 40, \, 80, \, 120, \, 200$ km s$^{-1}$ for
{\it IG1000, IG2000, IG4000} and {\it IG8000} (few Myr after the feedback
episode, again measured in the region $z<5$
kpc, $R<3$ kpc). The ISM motion dissipates to reach
$\sigma_{\rm hot}\sim 20$ km s$^{-1}$ at $t=50$ Myr for {\it IG1000} and
{\it IG2000}, and $\sigma_{\rm hot}\sim 60$ km s$^{-1}$ for {\it IG4000} and
{\it IG8000}.  As for the {\it CG} model, {\it IG} also shows a large velocity dispersion
for the off-center cold gas (run {\it IG2000}), with $\sigma_{\rm cold}
\sim 100$ km s$^{-1}$. Again, the main reason for this
discrepancy is the
different spatial regions probed by the the two gas phases. While for
the hot gas the whole off-center volume is used to compute
$\sigma_{\rm turb}$, the cold gas is present only in the smaller region $(R,z) \la
(0.8,2.5)$ kpc, where indeed the hot gas turbulence is stronger and
comparable to that of the cold gas ($\sim 100$ km s$^{-1}$).

The {\it IG} feedback simulations exhibit
similar charachteristics for the AGN-induced random motion as for the {\it
  CG-FB} models (Figure \ref{fig:vdisp}). The hot phase 3D velocity
dispersion varies from $\sim 50$ to $\sim 100$ km s$^{-1}$, while the
cold gas shows values about twice as those of the hot gas.

For the low mass galaxy {\it LM}, the chaotic ISM motion resulting
from the AGN activity is similar to 
that of the {\it IG-FB} models, and the 3D gas dispersion velocity
is shown in the bottom row of Figure
\ref{fig:vdisp}. We note that for the off-center cold gas of these {\it LM-FB}
models $\sigma_{\rm cold}$ is somewhat
lower with respect to the {\it CG-FB} and {\it IG-FB} runs, a
consequence of the lower free fall velocity of the {\it LM} galaxy.

For these isolated galaxies the velocity distribution of the cold gas
(Figure \ref{fig:histo}) is narrower than for the {\it CG}
model. For most of the time cold outflows are limited to $\approx 200$ km
s$^{-1}$, while the inflowing portion of the cold ISM commonly reaches $\sim 400$
km s$^{-1}$.

\section{Energetics}
\label{sec:energetics}

In this work we decided to be conservative regarding the strength of
the AGN feedback considered. We deliberately chose to adopt low power
outflows because they are expected to be the most common in normal
elliptical galaxies. For instance, even in the spectacular archetype of
AGN-ISM interaction NGC 5044, the power associated to the cavity system
is only $\la 6\times 10^{42}$ erg s$^{-1}$ \citep{david09}. More typical cavity
powers for massive ellipticals or groups range from $10^{41}$ to
$10^{42}$ erg s$^{-1}$ \citep{panagoulia2014}\footnote{It should be
  noted here that these estimates for low mass systems might be
  inaccurate, if very deep Chandra observations are not
  available. Furthermore, the fraction of outflow power dissipated in
  shocks is poorly known.}.

We do not show here the detailed evolution of the outflow energetics, but
limit ourself to a discussion of the basic properties of our feedback
simulations \citep[see][for a
detailed analysis of more sophisticated
models]{gasp12}. In agreement with the forementioned paper we find
that the total energy injected by the AGN slightly increases with
$v_{\rm jet}$ (although this trend reverts for {\it CG-FB8000}).
For models {\it CG-FB1000, CG-FB2000, CG-FB4000, CG-FB8000}
and {\it CG-FB16000} the total outflow energy $E_{\rm jet}^{\rm tot}$ 
delivered to the ISM at
$t=50$ Myr is $2.5\times 10^{55}, \; 7.0\times 10^{55}, \; 1.8 \times
10^{56}$, $7.2 \times 10^{55}$ and $6.0 \times 10^{56}$ erg, respectively. As found in
\citet{gasp12} we also note a definite trend for the duty cycle
(defined as the ratio between the time in which the outflow is active
to the total time of the simulation) to decrease with increasing
$v_{\rm jet}$: $2.3\times 10^{-2}$, $7.1\times 10^{-3}$, $1.9\times
10^{-3}$, $1.5\times 10^{-3}$ and $6.7\times 10^{-5}$ for the models listed above.
This is a signature of a self-regulating mechanism. In model {\it
  CG-FB16000} only a dozen AGN outbursts occur. 
The average outflow power, $P_{\rm jet} = E_{\rm jet}^{\rm tot}/($50
Myr $\times$ duty cycle$)$ is $6.9 \times 10^{41},\; 6.2 \times 10^{42}, \; 6.0
\times 10^{43}, \; 3.0\times 10^{43}$ and $ 6.7\times 10^{45}$ erg s$^{-1}$ for the five
considered models.
Single AGN outbursts show a typical 
power variation of $\sim 5$ with respect to the mean value.
Model {\it CG-FB16000} clearly shows a different feedback regime
with respect to the lower $v_{\rm jet}$ calculations. The AGN heating
operates through a sequence of relatively infrequent, very powerful
outbursts, separated by quescence periods lasting several Myr.
While this model is successful in reducing the total cooling rate down
to acceptable values, the central ISM often shows temperatures $\ga$ 2
keV, higher by a factor of $\sim 2$ than those in the center of real
galaxies/groups. The difficulty for the AGN heating to simultaneously
halt gas cooling and preserving low central temperatures is well known
and will not be discussed here \citep[see, e.g.][]{bm02,gasp12}.

For the isolated galaxy {\it IG}
the outflows energetics is somewhat different with respect of that of the {\it CG}
models. For instance, the total energy injected in the ISM is now $5.4\times
10^{55}$, $1.1 \times 10^{56}$ erg for model {\it IG-FB1000} and {\it
  IG-FB2000}, while the duty cycle is
 $3.3\times 10^{-3}$, $8.1 \times 10^{-4}$ and $5.9 \times 10^{-5}$ for
{\it IG-FB1000}, {\it IG-FB2000} and {\it IG-FB4000}.

\section{Discussion}
\label{sec:discuss}

\subsection{The Cold ISM of Elliptical galaxies}

Despite of the accuracy of the simulations presented here, we are well
aware that many results are subject to quantitative uncertainties. The
physics of the AGN feedback process still eludes our understanding and
has been represented only schematically. A source of error which is
difficult to evaluate in the simulation 
outcomes originates from the sometimes neglected numerical overcooling,
together with the intrinsic complexity of the cold-hot gas interaction
(which is impossible to reproduce in current galaxy-wide calculations).
Notwithstanding this cautionary note, we believe that a solid result
of our study (see also Brighenti, Mathews and Temi, 2015) is that AGN
activity triggers spatially extended gas cooling from the hot ISM
phase. This happens for a wide range of outflows parameters, a fact
which makes this prediction robust despite the uncertainties in the AGN
feedback process\footnote{Numerical experiments with different
  feedback methods, such as jets and cavities generated by cosmic rays
(Perrotta et al., in preparation), yield qualitatively similar results.}.
Of course, the net effect of the energy injected by the AGN is
to reduce (or suppress) the total cooling rate \citep[e.g.][]{gasp12}.

The off-center cold gas has an erratic distribution, following the cycle of the
AGN feedback, activated (by design) by the random presence of cold
(dropped-out) gas near the black hole
\citep{pizzolato2005,pizzolato2010,gaspari2013,gaspari2014}.
The exact amount of cold gas depends on the outflows parameters
(velocity, especially) and is therefore somewhat uncertain. In our
{\it CG} model (tailored on NGC $5044$) we find that $\sim 10^5 - 10^8$
M$_\odot$ of off-center cold gas are typically present, depending on
the feedback parameters. For the isolated galaxies {\it IG} and {\it
  LM} the amount of the cold gas is reduced, but still significant:
$\sim 10^4 - 10^6$ M$_\odot$. These values compare well with
observations \citep{macchetto1996,caon2000,david2014}. Especially for
the isolated systems, there are times when the cold gas is absent or
negligible (see Figure \ref{fig:mcold_feed}).
Indeed, that the mass budget of the cold ISM
phase undergoes unpredictable variations (cfr. Figure
\ref{fig:mcold_feed}), again in pace with the AGN activity.
It seems therefore plausible that AGN-induced cooling can explain, at
least in part, the large fraction
of intermediate/massive ellipticals with detectable H$\alpha$ emission
or even molecular gas, if $H_2$ and $CO$ can form fast enough in the
dust-poor cooled gas.

Our simulations cannot follow the evolution of the cooled gas with the
accuracy necessary to calculate what fraction of it ends up in
molecular form. $H_2$ can form by both grain surface and gas phase
reactions, with the dust grain processes dominant for gas metallicity
larger than $Z=10^{-4} - 10^{-5}$ solar \citep[e.g.][and references therein]{cazaux2009}.
The cold gas in our simulations has $\sim$ solar metallicity, but
the dust-to-gas ratio is $\delta \sim 10^{-5}$. This latter value is similar
to the dust-to-gas ratio of the model
with metallicity $Z=10^{-3}$ solar in \citet{cazaux2009}, which we can
therefore adopt to make order-of-magnitude considerations.
Using the
$H_2$ formation rates given in their Fig. 6, we find that the typical $H_2$ formation timescale
is $\approx 1$ ($50, 2000$) Myr for gas density $\sim 10^3$
($10^2,\, 10$) cm$^{-3}$. Unfortunately, for the highest quoted
densities, these numbers are close to
the lifetime of our simulated cold clouds, therefore no strong conclusions can be
drawn about the presence of molecular gas in the blobs. For $n=10$
cm$^{-3}$ no $H_2$ is expected to form. It is unclear which density
the gas cooled to $T\la 100$ K would reach --- our simulations do not have
the resolution to properly resolve the structure of the cold
clouds. Moreover it seems likely that magnetic or turbulent support would be
important in regulating the density profiles of the clouds.
Thus, in order to properly investigate this topic, a better calculation for the
thermal and dynamical evolution for the clouds is needed, coupled with
a $H_2$ (and $CO$) formation model which takes into account the chemical
properties of the cooled ISM and the appropriate dissociating UV
field. This is beyond the scope of the present paper, although such a
calculation would be crucial to properly interpret the molecular associations
detected in the central galaxy NGC 5044 by \citet{david2014}.

\citet{davis2011} found that most galaxies show kinematically
misaligned cold gas. This is usually interpreted as evidence that gas
has been accreted, through (minor or major) merging or from
smooth cosmological accretion. That is, that cold gas has an external
origin. Another possibility is the scenario
proposed by \citet{lagos2014}, where misaligned cold gas is generated internally
by cooling of a twisted hot gas halo. 
Undoubtedly these processes occur and may well explain the bulk of the
misaligned cases. However, it is interesting to note that AGN cooling
also predicts kinematically decoupled cold gas-stellar systems.
In fact, cold gas forms and moves preferentially along the outflow
axis, which may well be the rotation axis of the stellar body of the
galaxy. Therefore, looking to the system from a generic viewing angle,
stars and cold gas would indeed show a kinematic misalignment, even
though the gas origin is internal. Of course, the cold gas in this
simple scenario would likely show irregular appearance and kinematics --- 
smoothly rotating misaligned gas can be internally generated only
through the model described in \citet{lagos2014}.

Thus, AGN can play a double role in the galaxy life. It strongly
decreases the gas cooling process, solving the cooling flow problem
and quenching the star formation, causing the migration of the galaxy
to the red sequence. However, the dynamical interaction with AGN
outflows keeps the ISM ``alive'', giving rise to the episodic presence
of a significant cold phase and possibly triggering sporadic, weak
star formation events which could explain in part the recent SF
activity commonly detected in ETGs \citep{trager2000,kaviraj2007}.
While minor mergers are an attractive explanation
\citep{kaviraj2009},  we point out
that AGN-induced hot gas cooling is a natural, internal process which  
also potentially leads to low-level SF. 

\subsection{$H\alpha$ emission from warm clouds}
\label{sec:halpha}

Cold blobs, although they share similar properties, are made up of gas which has
cooled down to different temperatures ($1.8\la log \, T_{\rm blob} \la
4$).
Here we provide a rough estimate of the expected $H\alpha$ luminosity
of the spatially extended cold gas, which in turn depends on the
amount of the ionized cold ISM.
Column density of blobs, calculated as $N=\langle n \rangle
l$, where $\langle n \rangle$ is an average of the blob number density
and $l$ is its linear size, spans the range $3 \times 10^{19} \la N \la
3 \times 10^{21}$  cm$^{-2}$. 
Their number density is in the range
$0.4 \la n \la 50$ cm$^{-3}$ and their (smallest) characteristic size
is usually no more than few tens of parsec. 
This column density is usually larger than the maximum column
density $N_{\rm ion}$ that can be photoionized by the local ionizing
photon field, which includes the contributions of both the
stellar radiation (usually the dominant one at the clouds location) 
and the intergalactic background.

The mass fraction $M_{ion}$ of ionized gas in blobs has been estimated
in order to derive
the actual contribution of cooled gas emitting in $H\alpha$ band. Assuming
that blobs are spherical, the problem for a single blob reduces to an inverse
Str\"omgren sphere:
$\,4\: \pi {R_s}^2 \: \Delta R\: n^2 \: \alpha \:=\: 4\: \pi {R_s}^2 \:
n_{\gamma} \: c \:,$
where $R_s$ is the radius of the blob (or the smallest size of an elongated clump),
$\Delta R$ is the thickness of the actually ionized shell, $n$ and $n_{\gamma}$
the density of the gas inside the cool clump and the local ionizing photon density
respectively, $c$ the speed of light. Here, the number of recombinations
expected in the spherical shell has been equated to the number of ambient
incident photons which photoionize the outer layer of the blob; the recombination
coefficient is $\alpha \simeq 4 \times 10^{-13}$ cm$^3$ s$^{-1}$ \citep{spitzer78}.
The ionizing photon density $n_{\gamma}(r)$
has been estimated by following \cite{BM1997}, who considered an
elliptical galaxy of stellar mass $7.5 \times 10^{10}$ M$_\odot$ and
effective radius $1.72$ kpc, by scaling their $n_\gamma(r)$ for the
different stellar masses and effective radii of our galaxies.
The resulting ionizing photon density profile writes
$n_{\gamma}=A \,r_{\rm kpc}^{-1.5}$ cm$^{-3}$, where the constant
$A\,=\,10^{-3},\,3\times 10^{-3},\,2.4 \times 10^{-3}$ for {\it CG,
  IG} and {\it LM} respectively.

Solving for $\Delta R$, the thickness of the actually ionized layer of a typical
blob which is located at a certain distance from the center at a given time
and which has a certain density is estimated. From the value of
$\Delta R \approx (20,70,58) \,r_{\rm kpc}^{-1.5}/n^2$ pc (for models {\it
  CG, IG, LM}) it is possible to calculate the ionized mass of the blob:
\begin{equation}
M_{ion} \:\sim\: 4 \pi {R_s}^2 \: \Delta R \:\rho_{blob}\:\sim\: 3 M_{blob} \:
\frac{\Delta R}{R_s} \:\:\:,
\label{eq:m3}
\end{equation}

\noindent
where $\rho_{blob}$ is the density corresponding to the aforementioned $n$
and $R_s \simeq 30$ pc.
Unfortunately, we find that the fraction of ionized gas in a single clump varies
widely, depending on its density and location in the galaxy, spanning
the range $0.2 \%\: M_{blob} \le M_{ion} \le 50 \%\: M_{blob}$.

It is difficult and likely not worthy, given the uncertainties, 
to calculate the percentage of ionized mass of
every single blob. However, density and temperature of a large number of examined
clumps span the aforementioned ranges; the simulation time does not play a
crucial role in determining the density or the temperature of appearing (just
cooled) blobs; on the contrary, the evolution of blobs, when it can be followed,
shows that old blobs are denser than short-lived ones and that they are often
the result of the merging of ligther clumps. 

A further difficulty is that the simulated
blobs are under-pressurized with respect to the ambient gas, sometimes
by a factor of $10-100$ \citep[see also][]{li2014}. If temperatures of blobs are
thought to be reliable, the expected pressure equilibrium would
produce a corresponding increase in the density (neglecting magnetic
support)\footnote{Self-gravity, whose effect is not included in the code, cannot physically play
a crucial role in attempting to compress blobs and enhance their density,
the smaller size of a typical blob being usually far shorter than its Jeans length.}.
This, in turn, leads to a smaller linear size of the blob.
The combined effect of the increase of $\rho_{blob}$ and the drop in $l$ lowers
the mass fraction of ionized gas by a factor $\sim 5 \times 10^{-4}$ when a
typical blob of given mass is thought to have a two order of magnitude
larger density.
The aforementioned increase in $\rho_{blob}$ would lead to the enhancement
of dust growth in blobs, too, since $\dot{\rho}_{\:growth} \propto \rho$ (see
section \ref{sec:models}).
However, a cloud is expected to reach pressure equilibrium after a sound
crossing time $\tau_{sc}$; $\tau_{sc}$ spans the range $2-20$ Myr when
blobs of a typical size of $30$ pc and of different temperatures are considered.
The long $\tau_{sc}$ might be the responsible for the lack of pressure
equilibrium in coldest blobs, most blobs being younger than their $\tau_{sc}$
and more short living.
Observations indicate low thermal pressure in the ionized gas. 
For instance, in the central regions of M $87$ the [S II]$\lambda 6716/6731$
line ratio reveals electronic number densities ($n_e\simeq 25$ cm$^{-3}$) in
the emission line gas \citep{werner2013} which are lower than the ones obtained if pressure
equilibrium is supposed
($n_e\simeq 100$ cm$^{-3}$ in gas with $T\sim10^4$ K if a typical number
density of $n_{e,\,hot}\simeq 0.1$ cm$^{-3}$ in the $T_{hot}\sim10^7$ K
surrounding medium is adopted). 

These considerations suggest to not attempt a rigorous calculation of
the $H\alpha$ luminosity in our simulated galaxies. Nevertheless we
are able to give interesting upper limits in $L_{H\alpha}$ generated
by the warm non-nuclear gas. We rerun several models similar to those
described in the previous sections, but this time setting $T=10^4$ K as
lower limit in the gas temperature and assuming that all the cold gas
is ionized. In this way we can calculate firm upper limits for
$L_{H\alpha}$. For the {\it CG-FB} system we find for the off-center
gas an average value $L_{H\alpha,blob}\approx 5\times 10^{39}$ erg
s$^{-1}$, with typical variations of $\sim 5$ among the various {\it
  CG} models.
The $z$-axis filament can be much brighter, $L_{H\alpha,fil}\approx 10^{40}-10^{42}$ erg
s$^{-1}$, but most of its gas is indeed expected to be neutral or
molecular. The off-center warm gas in {\it IG} and {\it LM} models is
fainter than for the {\it CG}, an obvious result of the less
substantial cold ISM phase in these systems. The upper limits for the
{\it IG-FB} (when off-center cold gas is present --- see Figure
\ref{fig:mcold_feed}) vary in the range 
$L_{H\alpha,blob}\approx 10^{37} - 10^{39}$ erg s$^{-1}$, while {\it
  LM-FB} never exceeds $\sim 5\times 10^{37}$ erg s$^{-1}$. As before,
the upper limits for the filament can be one or two order of magnitude
brighter that those for off-center blobs.

\section{Conclusions}
\label{sec:conclusions}

In this paper we have presented hydrodynamical models of AGN outflows
expanding in the hot ISM of elliptical galaxies. We have designed
these numerical experiments to test the idea that AGN feedback, while
strongly quenching the total cooling rate, can nevertheless stimulate
some degree of hot gas cooling at a few kpc distance from the galaxy center
\citep[see
also][]{bm02,gasp12,gaspari2012b,Sharma12,McCourt12,costa2014}.
This process can generate a persistent cold ISM phase in early-type
galaxies. We explore the feasibility of this scenario in galaxies of
different mass and hot gas content, ranging from X-ray bright, group
centered ellipticals to galaxies of intermediate mass with a
relatively thin hot gas halo.

We have deliberately chosen a simple AGN feedback scheme. We assume
that AGN injects energy and momentum through non relativistic
outflows, activated by the gas cooling in the inner 100 pc \citep[see e.g.][]{gasp12}. We focus
here on relatively weak feedback, likely the most common regime in normal
galaxies.
The most significant results of our investigation can be summarized as
follow:

{\it i)} The interaction between AGN outflows and the ISM generates
non-linear density perturbations in regions of converging flows. These
perturbations trigger localized hot gas cooling at distances of several
kpc from the galaxy center. This confirms the early results of
\citet{bm02} for the {\it CG} model and extend them to
elliptical galaxies of lower mass ({\it IG} and {\it LM}
runs). \citet{li2014} have shown that this process occurs also in BCGs
at the center of massive clusters.

{\it ii)} The cooling process and the amount of off-center cooled gas are not
sensitive on the parameters of the outflows. This is a key result,
given the uncertainties of the AGN feedback mechanism and makes our
findings quite robust: AGN outbursts seem to stimulate large distance
cooling no matter what.

{\it iii)} Off-center cooling occurs when $t_{\rm cool}/t_{\rm ff} \la 70$
(in the unperturbed ISM),
but we do not expect that this is a strict threshold. Instead, the
spatially extended cooling process, no more closely related to thermal
instability when the inhomogeneities are strongly non-linear, depends on
the local (perturbation) value of $t_{\rm cool}/t_{\rm ff}$, that is
on the perturbation density contrast \citep[see
also][]{Joung2012}. Our calculations suggest that even moderate AGN
feedback is sufficient to generate density (entropy) inhomogeneities
which cool locally.

{\it iv)} The amount and the spatial distribution of the off-center cold ISM
is erratic (see Figure \ref{fig:mcold_feed}). For group-centered galaxies
(model {\it CG}), the extended cold phase has a typical mass of $10^6 -
10^7$ M$_\odot$, with time variation of a factor of $\sim 5$. 
For isolated galaxies ({\it IG} and
{\it LM}) periods of no off-center cold gas give turns to episodes of
significant cold and extended ISM presence, with mass $\sim 10^4 - 10^6$
M$_\odot$. A possible spin-off result concerns hot gas halos in massive
spiral galaxies. Given the similarity between detected hot coronae
around disk galaxies and the hot ISM in our LM
model, it appears conceivable that any outburst from the central black
hole (perhaps similar to the one responsible for the Fermi bubble)
could indeed trigger cooling of the hot corona, forming cold clouds
which will accrete on the galactic disk.

{\it v)} The cold gas initially shares the same kinematics of the hot out of
which it forms. Dispersion velocities in the range $150 - 250$ km
s$^{-1}$ for model {\it CG}, $100-200$ km s$^{-1}$ for {\it IG},
$50-100$ km s$^{-1}$ for {\it LM} are typical. This compares well with
measurements for the emission line component in ETGs
\citep[e.g.][]{caon2000}, while is generally lower than the values
given in \citet{werner2014} for [CII] emitting gas in a sample of
massive ellipticals. 

{\it vi)} Some gas condenses out from outflowing hot gas, resulting in a cold
outflow moving with characteristic velocity of a few 100s km s$^{-1}$
\citep[see also][]{costa2014}.
We also notice that cold gas moving mainly close and along the outflow
axis can appear kinematically misaligned with respect to the stars.

{\it vii)} We follow the evolution of the dust in both the hot and cold
gas. Dust is injected in the hot phase by the stellar mass loss, it
is destroyed by thermal sputtering and can grow in the cold clouds
by accretion of condensable elements. We find that the off-center cold
gas always has a very low dust-to-gas ratio, of the order of
$\delta_{\rm cold} \sim 10^{-5}$ for the {\it GC} and $\delta_{\rm
cold} \sim 10^{-3} - 10^{-4}$ for the lower mass systems.
Therefore, a strong dust deficiency is the distinctive sign for cold
gas originated internally from hot halo cooling. However, we
acknolewdge some degree of uncertainty for the latter conclusion. If
cold gas densities were grossly underestimated in our simulations (see
Section \ref{sec:halpha}) dust could grow fast enough to restore a
standard dust-to-gas ratio.

{\it viii)} It is difficult to predict the fraction of cold gas which
is ionized by the galactic ionizing radiation. However, upper limits
on the $H_\alpha$ radiation emitted by the off-center cold gas range
from $\sim 5 \times 10^{37}$ to $\sim 10^{40}$, with the higher
luminosity appropriate for the {\it CG} system.

{\it ix)} In the Appendix we show that the cold gas masses found in
our simulations are likely 
overestimated by a factor of a few. This is due to numerical
mixing and the following overcooling --- a phenomenon important when
small cold clouds move through a much hotter medium, sometimes
overlooked in the literature.

\section*{Acknowledgments}
FB is
supported in part by the Prin MIUR grant 2010LY5N2T ``The Chemical and
Dynamical Evolution of the Milky Way and Local Group Galaxies''. We
thank Bill Mathews, Pasquale Temi, Matteo Tomassetti and Filippo
Fraternali for useful discussions. The anonymous referee is thanked
for a prompt and constructive report.

\bibliographystyle{mn2e} \bibliography{cool_ref}

\appendix
\section{Numerical mixing and overcooling}

\begin{figure*}
\begin{center}
\adjincludegraphics[trim=2.6cm 6.2cm 2.cm 5.3cm, clip, max width=1.\textwidth]{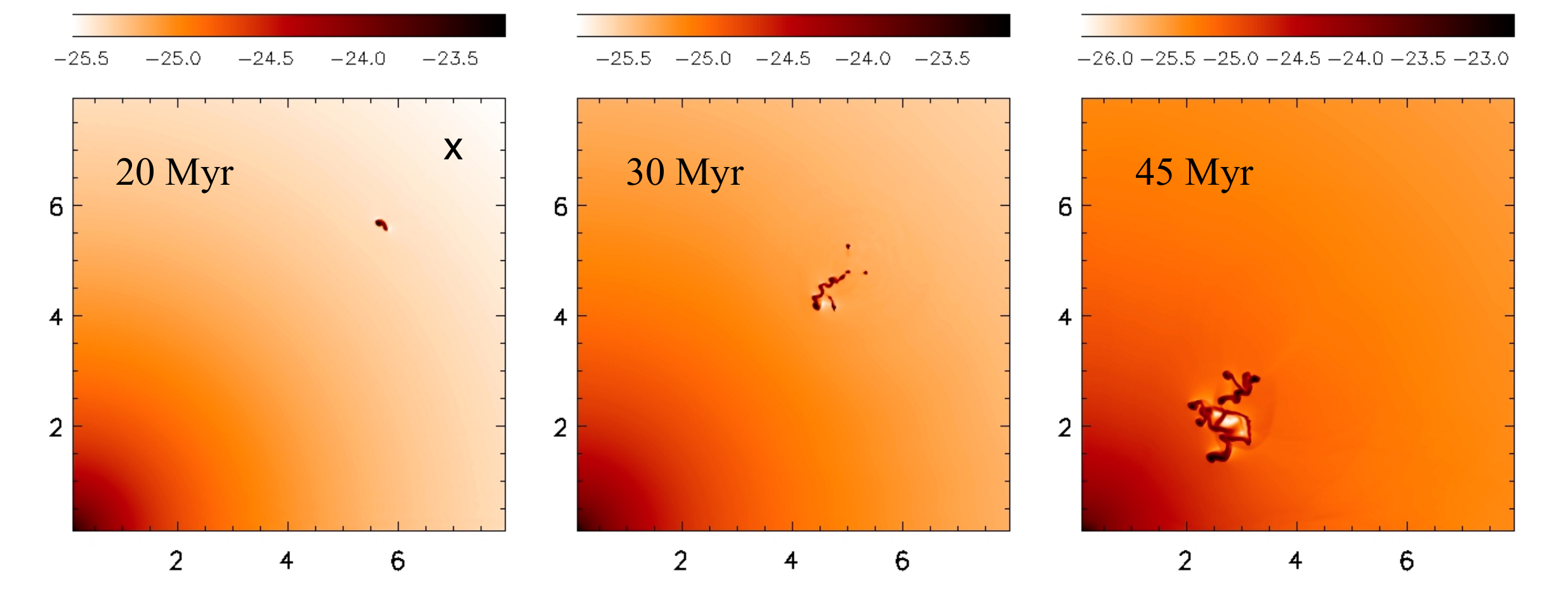} 
\end{center}
\caption{{\it Left column:} density maps for the
  ``one-zone'' experiment at three different times. In the left panel
  the ``x'' indicates the original location of the single cold zone. The {\it z-}axis is horizontal,
 the {\it R-}axis is vertical. Units are kpc. It
  is evident that the small cloud induces a ``snowball'' cooling
  effect as it moves through the grid.
Notice that the color scale differs for every panel.
}
\label{fig:onezone}
\end{figure*}

Numerical mixing at the interfaces between cold and hot gas can cause
large errors in the computed cooling rate --- not always recognized in
the literature. This is especially true
when the cold gas clouds are resolved in only a few zones and when the
difference in temperature and density between the two phases is
sizable. These are exactly the circumstances of our simulations, where
small blobs of $T\la 10^4$ K gas move through $\sim 10^7$ K hot
atmosphere. Numerical averaging at the surfaces separating cold and
hot gas results in zones of gas at intermediate density and
temperature. Radiative cooling can be either overestimated and
underestimated, depending on the flow parameters, with the former case
dominating in our simulations. 
Brighenti, Mathews \& Temi (2014) also
discuss the magnitude of this errors in simulations similar to those
presented here. They argue that adding a temperature dependent
``dropout'' term to the continuity equation, which removes from the
grid the currently cooling gas, largely reduces the
spurious overcooling. The trade off of this procedure is, of course,
the inability to follow the subsequent dynamics of the cooled gas; only
its amount and the location of its formation can be estimated. This
term has already been introduced in Section 2.2.2, see Brighenti,
Mathews \& Temi (2014) for more details. The term added to the
continuity equation reads:
\begin{equation}
\left (\frac{\partial \rho}{\partial t} \right )_{\rm do}=
-q(T)\frac{\rho}{t_{\rm cool}}; \;\;\;\; q(T)=q_0\exp{[-(T/T_c)^2]}\;\;
\label{eq:app1}
\end{equation}
and a corresponding term is needed to the energy equation (see
Brighenti, Mathews \& Temi 2014). Here we choose $T_c=5\times 10^5$ K.
These dropout terms are negligibly for $T\ga
T_c$. By targeting only gas with $T\la T_c$ means that we remove gas
that is currently cooling and will reach $T\la 10^4$ K in a very short
time. That is why this method produces accurate cooling (dropping out)
rates, providing $T_c\la 10^6$ K \citep[see][]{bm02}. The precise value
of $q_0$ hardly matters and we choose $q_0=2$ to match our previous
calculations. 

Needless to say, this
technique to alleviate numerical overcooling is not perfect. A sensible way
to gauge the magnitude of this error is to calculate similar models
with and without the dropout term (acting in the whole numerical
grid), and compare the amount of cooled
(dropped out) gas in the two cases; the physical truth lies in between
these two extrema, with the dropout value likely to be more accurate.

\subsection{``One cold zone'' experiment}

Before to make this comparison for the {\it CG} model, we show with a
suite of idealized, somewhat pedantic, numerical experiments 
the occurrence and the effect of the overcooling. 
We follow the evolution of a one-zone cold cloud, located
in the same gas atmosphere of the {\it CG} model at $(R,z)=(7,7)$ kpc.
The temperature of the zone is set to $10^4$ K and its density is
calculated assuming pressure equilibrium with the surrounding hot
gas. The numerical resolutions adopted for the calculations are $\Delta R = \Delta
z = 20 \;\; {\rm or} \;\; 10$ pc. 
The dropout term above is not used in this experiment, and
the cold gas is simply advected under gravity toward the galaxy
center. The cloud evolution for the radiative case, resolution 20 pc, 
is shown in Figure \ref{fig:onezone}. 

\begin{figure}
\begin{center}
\adjincludegraphics[trim=3.25cm 8.5cm 5.5cm 11.cm, clip, max width=0.52\textwidth]{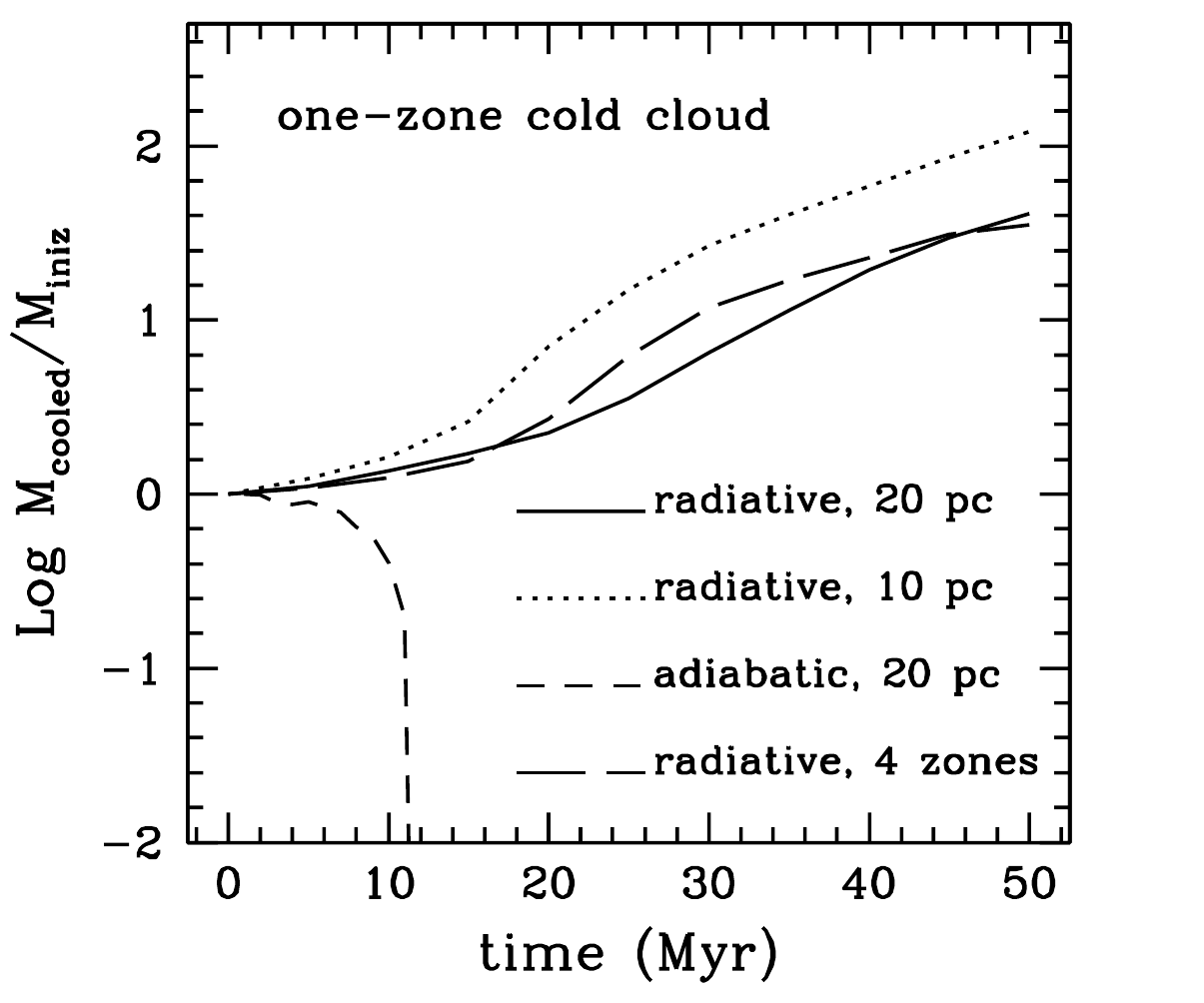} 
\end{center}
\caption{Time evolution of the cold gas mass for the one-zone (and
  one four-zones) experiments. For every run the mass is normalized to
  the initial cold mass. In radiative simulations numerical
  mixing causes overcooling, while in the adiabatic calculation
  the spurious mixing erases the cloud in a short time.
}
\label{fig:mcold_onezone}
\end{figure}

At $t=20$ Myr the mass of the cloud has grown by a factor 2.4, as can
be seen in Figure \ref{fig:mcold_onezone} (solid line), where we show
the cold gas mass evolution (normalized to the initial cold mass). 
This increase in the cold gas mass is spurious and it is due to
numerical overcooling. 
While this mimics some physical
processes (for example the condensation for clouds larger than the
Field length, when thermal conduction is present), we stress that in
the hydrodynamical equations solved no transport terms (which deal
with microscopic fenomena such as diffusion) are present,
therefore any increment of the cold gas mass must be regarded as
spurious. At the end of the calculation (50 Myr), the cold gas has
grown by a factor $\ga 40$.

If the radiative cooling is turned off, numerical mixing consumes the
cold cloud, in a time $t\sim 10$ Myr (Figure \ref{fig:mcold_onezone},
dashed line). After this time the original cloud gas has been mixed
with hot gas and raised its temperature to values $> 5\times 10^5$ K,
our threshold for the definition of cold gas.

We run two more simulations at higher resolution (10 pc). The first
shows that a (radiative) single zone cloud increases its mass by a
factor $\ga 100$ (Figure \ref{fig:mcold_onezone}, dotted line). 
In the second, high resolution experiment we
consider a 4-zones cloud (which is then physically identical to the
single-zone cloud at 20 pc resolution). In this case, illustrated in 
Figure \ref{fig:mcold_onezone} with a long-dashed line, the cold gas
mass growth is almost identical to the model described by the solid
line. This implies that a modest increase in the numerical resolution
likely does not solve the numerical mixing problem.

These few simple calculations, although extreme, are relevant for our
calculations. It is our experience that regions of cold ($\sim 10^4$
K) gas only one or a few zones across can arise naturally in radiating
gas dynamics \citep{koyama2004}. Thus, small elements of cold gas can trigger large
unintended numerical overcooling due to unphysical mixing with much
hotter gas. Conversely, in very slowly cooling or adiabatic flows,
some degree of numerical overheating can occur. The physical truth
lies between these two extremes.
\subsection{Numerical overcooling in the {\it CG-FB} models}

\begin{figure}
\begin{center}
\adjincludegraphics[trim=5cm 10.6cm 5.7cm 10.9cm, clip, max width=0.52\textwidth]{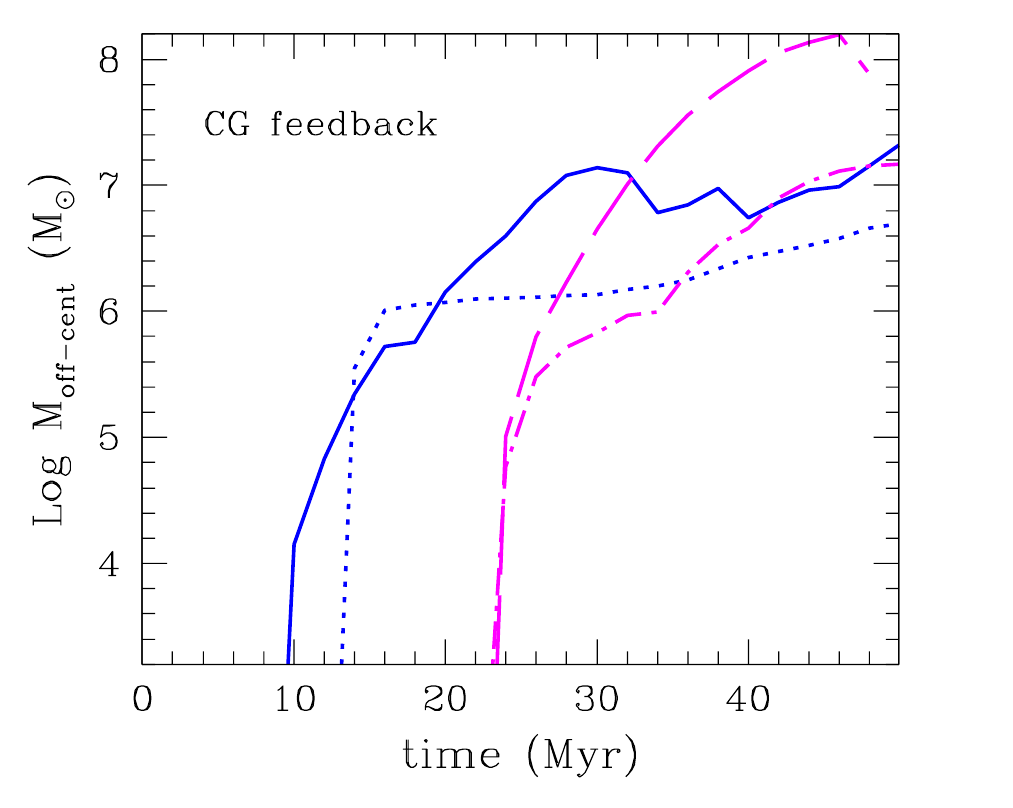} 
\end{center}
\caption{Time evolution of the cold gas mass for two {\it CG-FB}
  models. The solid blue line indicates the same {\it CG-FB4000} model
  shown in Figure \ref{fig:mcold_feed}. In this calculation no
  dropout term has been used. The blue dotted line represents the
  same model, but with the term \ref{eq:app1} adopted. The two magenta
  lines show the standard (dashed line) and the dropout (dot-dashed)
  version of {\it CG-FB16000}.
}
\label{fig:mcold_do}
\end{figure}

In this section we address the effect of numerical overcooling in some {\it
  CG-FB} models. We also performed experiments for {\it IG-FB}
calculations, with similar results. For the sake of conciseness, we
present results for only two runs of {\it CG-FB}
with jet velocities $v_{\rm jet} = 4000$ and $16000$ km s$^{-1}$.
These calculations are identical to those described in Section
\ref{sec:cgfeed}, but the term in equation \ref{eq:app1} (and an analogue term in
the energy equation) is activated on the whole numerical grid. 
In Figure \ref{fig:mcold_do} we show the cumulative mass dropped out
in the off-center region and compare it with the cold gas mass found in
the same region for the respective standard models (see Section
\ref{sec:cgfeed}). The masses which are portrayed in Figure \ref{fig:mcold_do}
have different meaning. For the standard models, the displayed
mass represents the instantaneous cold gas mass present in the off-center
region: this cold gas then flows toward the center, disappearing from
that region. For the dropout runs it is instead the total gas mass
which has been cooled and removed from the grid since $t=0$. 
That is why it is a monotonically increasing value.

The effect of the overcooling in the case of no-dropout is clear, especially for the $v_{\rm
  jet}=16000$ km s$^{-1}$ model. In the temperature map, not shown
  here, it is evident as a large number of small cold blobs form at
  $t\sim 25$ Myr for $(R, z) \approx (1.5, 4)$ kpc. These blobs,
  initially only a few tens pc in size, grow in size and mass by
  numerical mixing as they move through the grid, exactly as
  depicted in Figure \ref{fig:onezone}. That is, overcooling has
  been triggered. The mass discrepancy can be as large as one order of
  magnitude, and will be not reduced by increasing the numerical
  resolution by a factor of a few. Therefore, the cold gas mass values
  quoted in the previous sections are uncertain by (at least) a factor
  of a few. Although we argue that the dropout masses are closer
  to the physical truth than the cold gas masses calculated in the
  standard fashion, we focused on the latter estimates in the text, mostly for
  the sake of tradition. However, we stress the need to evaluate, whenever
  possible, the uncertainty in the cooling gas mass due to numerical
  overcooling, using a drop out procedure or other appropriate experiments.

\label{lastpage}
\end{document}